\documentclass[preprint,onecolumn,nofootinbib,prd]{revtex4}
\pdfoutput=1
\usepackage{amsfonts}
\usepackage{amssymb}
\usepackage{graphicx}
\usepackage{color}
\usepackage{epsfig}
\usepackage{remreset}
\usepackage{amsmath,amsthm,ulem}
\usepackage{mathrsfs}
\usepackage{color}
\usepackage{float}
\usepackage{tensor}
\usepackage[colorlinks=true,linkcolor=blue,urlcolor=blue,filecolor=black,citecolor=red,pdfstartview=FitV,
pdftitle={},pdfsubject={},pdfkeywords={},pdfpagemode=None,bookmarksopen=true]{hyperref}
\begin{document}
\baselineskip=0.5 cm
\title{Horizon curvature and spacetime structure influences on black hole scalarization}

\author{Hong Guo$ ^{1,2}$}
\email{gh710105@gmail.com}
\author{Xiao-Mei Kuang$ ^{1}$}
\email{xmeikuang@yzu.edu.cn (corresponding author)}
\author{Eleftherios Papantonopoulos$ ^{3}$}
\email{lpapa@central.ntua.gr}
\author{Bin Wang$ ^{2,1}$}
\email{wang$_$b@sjtu.edu.cn}

\affiliation{$^1$ Center for Gravitation and Cosmology, College of Physical Science and Technology, Yangzhou University, Yangzhou 225009, China}
\affiliation{$^2$ School of Aeronautics and Astronautics, Shanghai Jiao Tong University, Shanghai 200240, China}
\affiliation{$^3$ Physics Division,
National Technical University of Athens, 15780 Zografou Campus,
Athens, Greece.}

\date{\today}

\begin{abstract}
Black hole spontaneous scalarization  has been attracting more and more attention  as it circumvents the well-known no-hair theorems. In this work, we  study the scalarization in Einstein-scalar-Gauss-Bonnet theory with a probe scalar field in a black hole background with different curvatures. We first probe the signal of black hole scalarization with positive curvature in different spacetimes. The scalar field in AdS spacetime could be formed easier than that in flat case.
Then, we investigate the scalar field around AdS black holes with negative and zero curvatures. Comparing with negative and zero cases, the scalar field near AdS black hole with positive curvature could be much easier to emerge. And in negative curvature case, the scalar field is the most difficult to be bounded near the horizon.
\end{abstract}

\maketitle

\tableofcontents

\section{Introduction}
Experimental progress on gravitational waves \cite{GW1,GW2,GW3} and the shadow of the $M87$ black hole \cite{EHT} further demonstrates the great success of Einstein's general relativity (GR). Yet it is unabated that the  GR theory should be generalized, and in the generalized theories extra fields or higher curvature terms are always involved in the action \cite{Nojiri:2006ri,Clifton:2011jh,MG3}. So physicists have proposed various modified gravitational theories which indeed provide richer framework and significantly help us further understand GR as well as our universe.

Among them, the scalar-tensor theories which introduces a scalar field into the action attract lots of attention \cite{Fujii}. When the scalar field backreacts to the background metric, one could  expect that hairy black hole solutions would be generated. One of the first hairy black hole solution in an asymptotically flat spacetime was discussed in \cite{BBMB} but soon the solution was argued to be unstable because the scalar field is divergent on the event horizon \cite{bronnikov}. However, such irregular behaviour of the scalar field  on the horizon was avoided in asymptotically AdS/dS spacetime with a presence of  a cosmological constant in the gravity theory. It was found that the resulting hairy black hole solutions have a regular scalar field behaviour and all the possible divergence could be hidden behind the horizon \cite{Martinez:1996gn,Banados:1992wn,Martinez:2004nb,Zloshchastiev:2004ny,Martinez:2002ru,Dotti:2007cp,Torii:2001pg,Winstanley:2002jt,Martinez:2006an,Kolyvaris:2009pc,Charmousis:2014zaa,Khodadi:2020jij}.

The effects of higher-order curvature terms become more significant as we are exploring the strong field regime of gravity via detections of gravitational waves and black hole shadows. It is known that the inclusion of such terms probably bring in the well-known ghost problem \cite{stelle}, and Gauss-Bonnet (GB) corrections is a counter-case which  is ghost-free. But it becomes a topological term in four-dimensional spacetime and has no dynamics in field equations when it is minimally coupled with Einstein-Hilbert action.  A way to make this term meaningful in four-dimensional spacetimes is to consider its coupling with a scalar field \cite{stringT}. As a special scalar-tensor theory with higher derivatives \cite{Horndeski,Deffayet:2009mn}, the Einstein-scalar-Gauss-Bonnet (ESGB) gravity recently has attracted a lot of attention. Especially, the introduction of this coupling could lead to hairy black holes.  Various black hole solutions and compact objects in the four-dimensional ESGB theories were studied in the literature \cite{Mignemi_1993,Kanti_1996,Torrii_1996,Ayzenberg_2014,Kleihaus_2011,Kleihaus_2016a}.

Recently, the spontaneous scalarization with particular coupling function in ESGB theory is widely investigated. In this setup, besides GR solutions with a trivial scalar field configuration, the scalarized hairy solutions for black holes and stars could also exist, which evades the no-hair theorems \cite{Bekenstein,Antoniou_2018,Antoniou:2017hxj}. It was shown in \cite{Doneva:2017bvd} that below a certain mass   the Schwarzschild black hole background may become unstable in regions of strong curvature, and then when the scalar field backreacts to the metric,  a scalarized hairy black hole emerges and it is physically favorable. A natural extension of these results was to consider the case of nonzero black hole charge. A large set of coupling functions between the Gauss-Bonnet invariant and the scalar field was considered  in order to understand better the behaviour of the scalarized solutions  \cite{Doneva:2018rou}-\cite{Fernandes:2019rez} .

Various generalizations of the spontaneous scalarization procedure were followed in the literature. The scalarization due to a coupling of a scalar field to Ricci scalar was studied in \cite{Herdeiro:2019yjy} and in the presence of Chern-Simons invariant was studied in \cite{Brihaye:2018bgc}. The generalized study of scalarized black hole solutions and compact objects in asymptotical flat spacetime of ESGB theories was discussed in \cite{Silva:2017uqg}-\cite{Hunter:2020wkd}.  The spontaneous scalarization of asymptotically AdS/dS black holes in  ESGB theory with a negative/positive cosmological constant was extended in \cite{Bakopoulos:2018nui,Brihaye:2019gla,Bakopoulos:2019tvc,Bakopoulos:2020dfg,Lin:2020asf}. Especially, the connections of asymptotically AdS  black holes  scalarization  with  holographic phase transitions in the dual boundary theory was studied in \cite{Brihaye:2019dck,Guo:2020sdu}. Recently the spontaneous scalarization in f(R) gravity theories in the presence of a scalar field minimally coupled to gravity with a self-interacting potential was discussed in \cite{Tang:2020sjs}. Also in ESGB theory the spin-induced black hole spontaneous scalarization, which is the outcome of linear tachyonic instability triggered by rapid rotation, was explored in \cite{Collodel:2019kkx,Dima:2020yac,Doneva:2020kfv,Herdeiro:2020wei,Berti:2020kgk,Zhang:2020pko}.

The scalarization in asymptotical flat, AdS and dS black holes in ESGB gravity theory was studied, however detailed information on how different spacetime structures influence  the scalarization process is not disclosed. An interesting question is how the cosmological constant leaves its imprint on the black hole scalarization and in which kind of spacetime the scalar field could be formed more easily. Besides, the studies of spontaneous scalarization in the existed literatures were focused on the black hole with spherical horizon, but as known in AdS spacetime, one can have AdS black holes with different topologies and it is interesting to study how these topologies of the black hole horizon affect the scalarization procedure\footnote{Note that in \cite{Motohashi:2018mql}, the authors  discussed the influence of the topology in the matter configuration on the spontaneous scalarization. They found that comparing to the spherical distribution, the planar symmetric matter distribution makes the spontaneous scalarization easier to occur.}.
Thus, the aim of this work is to probe the signal of scalarization in various  black holes with different horizon curvature and try to understand its effects on the scalarization processes. To this end, we first consider the test scalar field perturbation and analyze its effective potential in various black hole backgrounds. Then we study the tachyonic instabilities of the scalar field near the horizon by solving the perturbation equation.
We find that the scalar hair around the black hole with spherical horizon is easier to form in AdS spacetime than in flat case.  While in dS case, the regular condition near cosmological horizon leads to divergence at large distance, so that the test field approximation breaks down. Moreover, our study on the effect of horizon curvatures indicates that the scalar hair around AdS black hole with toroidal horizon could be easier to be formed than that with hyperbolic horizon, but more difficult than that with spherical horizon. Both the effective potential analysis and the test field solution conclude the above observations.

The work is organized as follows. We write down the general equation of motion of a test scalar field  in the background of  four dimensional  black holes with different topologies in ESGB gravity theory in section \ref{ba}. In section \ref{sec:k=1}, we study the signal of scalarization of a black hole with positive curvature in different spacetimes. Then we investigate the signal of a black hole scalarization with negative curvature and zero curvature in section \ref{sec:k=-1} and section \ref{sec:k=0} respectively. The last section is devoted to conclusions and discussions.

\section{Scalar Field Perturbations in  Einstein-Scalar-Gauss-Bonnet Theories}
\label{ba}

We consider four-dimensional ESGB theories with the action
\begin{eqnarray}
S=\frac{1}{16\pi G}\int d^4x \sqrt{-g}
\Big[R-2\Lambda- \nabla_\mu \Phi \nabla^\mu \Phi  - m^2\Phi^2+ \lambda^2 f(\Phi) {\cal R}^2_{GB} \Big]~,\label{eq:quadratic}
\end{eqnarray}
where $\Lambda$ is the cosmological constant, $\lambda$ is the  coupling constant between Gauss-Bonnet term and scalar field with the dimension of length and $\nabla_{\mu}$ denotes the covariant derivative. The scalar field coupling function $f(\Phi)$ only depends on the neutral scalar field $\Phi$, and the Gauss-Bonnet term is given by
\begin{equation}
{\cal R}^2_{GB}=R^2 - 4 R_{\mu\nu} R^{\mu\nu} + R_{\mu\nu\alpha\beta}R^{\mu\nu\alpha\beta}~,
\end{equation}
where $R, R_{\mu\nu}$ and $R_{\mu\nu\alpha\beta}$ are the Ricci scalar, Ricci and Riemann tensors, respectively.
The Einstein equation and Klein-Gordon equation can be derived from the above action  as
\begin{eqnarray}\label{FE}
&G_{\mu\nu}-\Lambda g_{\mu\nu} + \Gamma_{\mu\nu}= \nabla_\mu\Phi \nabla_\nu\Phi - \frac{1}{2}g_{\mu\nu} \nabla_\alpha\Phi \nabla^\alpha\Phi -\frac{1}{2}g_{\mu\nu}m^2\Phi^2~, \\
&\frac{1}{\sqrt{-g}}\partial_\mu(\sqrt{-g}\partial^\mu)\Phi=m^2\Phi - \frac{\lambda^2}{2} \frac{df(\Phi)}{d\Phi} {\cal R}^2_{GB}~\label{FE2},
\end{eqnarray}
where the energy-momentum tensor $\Gamma_{\mu\nu}$ is given  by
\begin{eqnarray}
\Gamma_{\mu\nu}&=& - R(\nabla_\mu\Psi_{\nu} + \nabla_\nu\Psi_{\mu} ) - 4\nabla^\alpha\Psi_{\alpha}\left(R_{\mu\nu} - \frac{1}{2}R g_{\mu\nu}\right) +
4R_{\mu\alpha}\nabla^\alpha\Psi_{\nu} + 4R_{\nu\alpha}\nabla^\alpha\Psi_{\mu} \nonumber \\
&& - 4 g_{\mu\nu} R^{\alpha\beta}\nabla_\alpha\Psi_{\beta}
 + \,  4 R^{\beta}_{\;\mu\alpha\nu}\nabla^\alpha\Psi_{\beta}~,
\end{eqnarray}
with $\Psi_{\mu}= \lambda^2 \frac{df(\Phi)}{d\Phi}\nabla_\mu\Phi$.

It is obvious that different forms of the coupling function $f(\Phi)$ shall give different properties of the ESGB theory. As addressed in  \cite{Doneva:2017bvd}, $f(\Phi)$ could satisfy  conditions $\frac{df(\Phi)}{d\Phi}\mid_{\Phi=0}=0$ and $\frac{d^2f(\Phi)}{d\Phi^2}\mid_{\Phi=0}=b^2>0$, where $b$ is a constant, in order to admit Schwarzschild black hole as a background solution and further explore the black hole scalarization in the ESGB theory. Moreover, one usually assumes that the scalar field vanishes at infinity and normalizes the constant $b$ to  unity. Thus, in this work, we shall choose the form of the coupling function as $f(\Phi)=\frac{1}{2}(1-e^{-\Phi^2})$ which also satisfies $f(\Phi=0)=0$.

When the scalar field vanishes, i.e., $\Phi=0$, the gravity theory admits  black hole solutions, whose metric takes the general form,
\begin{equation}\label{metric}
ds^2=-g(r)dt^2+\frac{dr^2}{g(r)}+r^2d\overrightarrow{x}_{k,2}^2~,~~~~g(r)=k-\frac{M}{r}-\frac{\Lambda r^2}{3}~,
\end{equation}
where $M$ is the integral constant related the black hole mass, and $k$ determines the curvature of two dimensional geometry near the horizon with the line element $d\overrightarrow{x}_{k,2}^2$ given by
\begin{equation}
d\overrightarrow{x}_{k,2}^2=\left\lbrace\begin{matrix}
d\Omega^2_{2}&=&d\theta^2+\sin^2\theta\, d\psi^2\ \ \ &\ \  {\rm for\ }k=+1\ \,,\cr
d\ell^2_{2}&=& dx^2+dy^2\ \ \qquad
\qquad&\ \ {\rm for\ }k=0\ \ \,,\cr
d\sigma^2_{2}&=&d\theta^2+\sinh^2\theta\, d\psi^2&\ \  {\rm for\ }k=-1\,,
\end{matrix}\right. \label{geometries}
\end{equation}
where $k=1,0$ and $-1$ correspond to black hole horizon with spherical, planar and hyperbolic topology, respectively.
Then the corresponding GB term can be easily computed as
\begin{equation}\label{GB}
{\cal R}^2_{GB}=\frac{4}{r^2}\left[g'(r)^2+(g(r)-k)g''(r)\right]~.
\end{equation}
In order to explore the (in)stability of the background black holes, we consider  a small fluctuation in the form $\delta \Phi=e^{-i\omega t}\phi(r)Y(x_1,x_2)$ where $Y(x_1,x_2)$ satisfies  Laplace-Beltrami equation $\Box_2Y(x_1,x_2)=-AY(x_1,x_2)$. We consider that
$Y(x_1,x_2)$ is spherical harmonics $Y_{\ell, m}$ for positive curvature so that $A=\ell(\ell+1)$; For zero curvature $Y(x_1,x_2)$ is plane wave case as $e^{i\mathcal{K}(x+y)}$ so that $A=\mathcal{K}^2$; and for negative curvature $Y(x_1,x_2)$ is spherical harmonics  $Y_{-\frac{1}{2}\pm i\zeta,\mathfrak{m}}$ and $A=\frac{1}{4}+\zeta^2$ \cite{Koutsoumbas:2008pw,Wang:2001tk}. Then we use the tortoise coordinate $dr_*=\frac{dr}{g(r)}$, and bring the Klein-Gordon equation into a Schrodinger-like form by  reforming $\phi(r)=\frac{\varphi(r)}{r}$
\begin{eqnarray}\label{varphi}
\frac{\partial^2 \varphi(r)}{\partial r_{*}^2}+(\omega^2-V(r)) \varphi(r)=0~,
\end{eqnarray}
where the effective potential takes the form
\begin{equation}\label{eq-potential}
	V(r)=g(r)\left(\frac{g'(r)}{r}+\frac{A}{r^2}+m^2-\frac{\lambda^2}{2}{\cal R}^2_{GB} \right).
\end{equation}
Noted that in the Schrodinger-like form \eqref{varphi} and \eqref{eq-potential} of the perturbation, we have only kept the leading term of the expansion $\frac{df(\Phi)}{d\Phi}\simeq\Phi(1-\Phi^2+\frac{1}{2}\Phi^4+\mathcal{O}(\Phi^2))$ in the potential function.

We then analyze the signal of scalarization of the system by computing the effective potential with the lowest mode $(\ell=\mathcal{K}=\zeta=0)$. For simplification, we could focus on the static perturbation with $\omega=0$, thereby we shall solve the radial perturbed equation for $\phi(r)$,
\begin{eqnarray}\label{scalareq}
\phi''(r)+\left(\frac{2}{r}+\frac{g'(r)}{g(r)}\right)\phi'(r)-\left(\frac{A}{r^2g(r)}+ \frac{m^2}{g(r)}\right)\phi(r)+\frac{\lambda^2}{2g(r)}{\cal R}^2_{GB}\frac{df(\phi)}{d\phi}= 0~,
\end{eqnarray}
and see how the scalar field with the lowest mode emerges as the Gauss-Bonnet coupling increases. To control the variables of the black holes and compare the results, we shall set $r_h=1$ in various cases, namely, we fix the size of the black holes such that the black hole mass is solved via $g(r=r_h)=g(r_h=1)=0$ in the following discussion. We would numerically employ the spectral method to solve the differential equations.


Since the existence of above black hole with different curvatures is dependent on the type of asymptotic background, i.e., the sign of $\Lambda$, we would discuss the signal of scalarization of various black holes in terms of the curvature. Moreover, we shall rewrite $\Lambda=\pm3/L^2$ where $L$ is the curvature AdS radius.
As it was discussed in \cite{Guo:2020sdu,Kolyvaris:2013zfa} because the scalar field is coupled to GB invariant, the effective mass of the scalar field receives a contribution from this coupling. Then for various values of the coupling, the effective mass of the scalar field becomes  tachyonic outside the background black hole and the fluctuations are unstable destabilizing in this way the background metric.


\section{Signal of Scalarization of Black Holes with Positive Curvature}

\label{sec:k=1}
In this section, we focus on the background solution with positive curvature, i.e., $k=1$ in \eqref{metric}, in which case there exist asymptotical flat $(\Lambda=0)$, AdS $(\Lambda<0)$ and dS $(\Lambda>0)$ black holes as possible backgrounds.  Consequently, we shall separately  study the signal of scalarization in all these asymptotical backgrounds, and then do comparisons on the emergence of non-trivial scalar field which is introduced by the Gauss-Bonnet coupling in these different spacetime backgrounds.

\subsection{Asymptotical flat black hole}

For $\Lambda=0$, the asymptotical flat black hole (AFBH) with spherical horizon ($k=1$) is just the Schwarzschild black hole. In this case, the redshift function in the field equation \eqref{scalareq} is $g(r)=1-\frac{M}{r}$ where $M=r_h=1$ and ${\cal R}^2_{GB}=\frac{4}{r^2}(g'(r)^2+(g(r)-1)g''(r))$.

In this case, the potential \eqref{eq-potential} for the lowest mode $\ell=0$ of the massless scalar field $(m=0)$ is
 \begin{equation}
	V(r)=(1-\frac{M}{r})\left(\frac{M}{r^3}-\frac{6\lambda^2M^2}{r^6}\right)=(1-\frac{1}{r})\left(\frac{1}{r^3}-\frac{6\lambda^2}{r^6}\right)
\end{equation}
where we have inserted $M=1$ in the second equality.
A sufficient condition for the existence of an unstable scalar field mode is \cite{Buell,Doneva:2017bvd}
\begin{equation}\label{condition}
	\int^{\infty}_{-\infty}V(r_*)dr_*=\int^{\infty}_{M}\frac{V(r)}{g(r)}dr<0,
\end{equation}
which leads to $\lambda^2>\frac{5}{12}M^2=\frac{5}{12}$. This indicates that the tachyonic instability emerges when the coupling constant satisfies $\lambda^2>\frac{5}{12}$. Especially, near the horizon $r\to 1$, we can reduce from equation \eqref{scalareq} that the scalar field $\phi(1)$ and its derivative  $\phi'(1)$ are both regular with a relation
\begin{eqnarray}
\phi'(1)+6\lambda^2\phi(1)e^{-\phi(1)^2}=0~.
\end{eqnarray}
It is obvious that the contribution of the GB coupling could control the near horizon behaviour of the scalar field. While at infinity $r\to +\infty$, the scalar field behaves as
\begin{equation}\label{ByFlat}
\phi(r\to \infty)\to \phi_\infty+\frac{\phi_-}{r}+\frac{\phi_+}{r^2}+\mathcal{O}(\frac{1}{r^3}).
\end{equation}
We would set $\phi_\infty=0$ as the Dirichlet boundary condition so that the scalar field converges to zero as $r\to \infty$ which is well behaved. Thus, we can consider the scalar field as a probed field to study the signal of the scalarization.  Also $\phi_-$  could be treated as the scalar charge.

\begin{figure}[thbp]
\center{
  \includegraphics[scale=0.7]{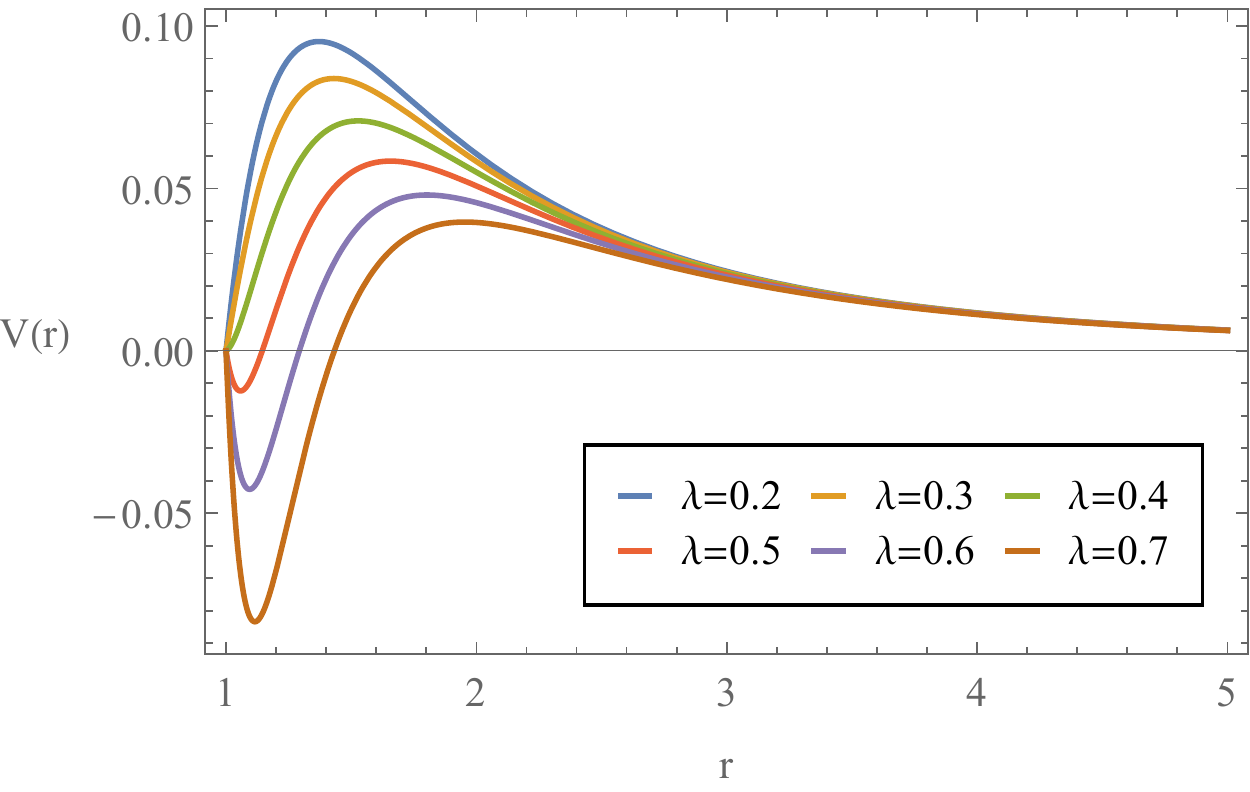}
  \caption{The effective potential of the scalar field in asymptotical flat  schwarzschild black hole with different values of $\lambda$.}\label{scapoten}}
\end{figure}

In Fig. \ref{scapoten}, we show the effective potential of the scalar field as the function of radial coordinate.
When $\lambda=0$ without GB coupling, there is a potential barrier. As the GB coupling is turned on and increased, the barrier is suppressed, and a negative potential well form near the horizon which could trigger an  instability and then lead to the development of scalar hair.

\begin{figure}[thbp]
\center{
  \includegraphics[scale=0.42]{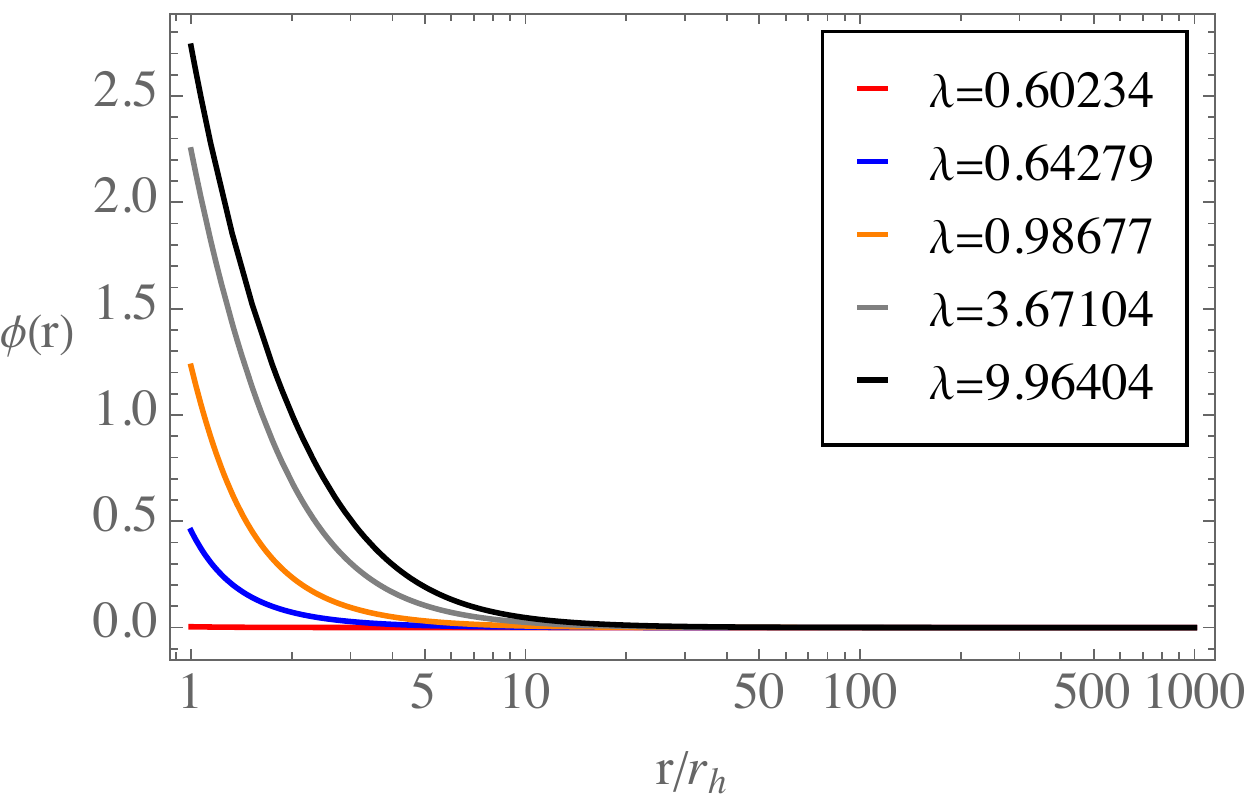}
  \includegraphics[scale=0.42]{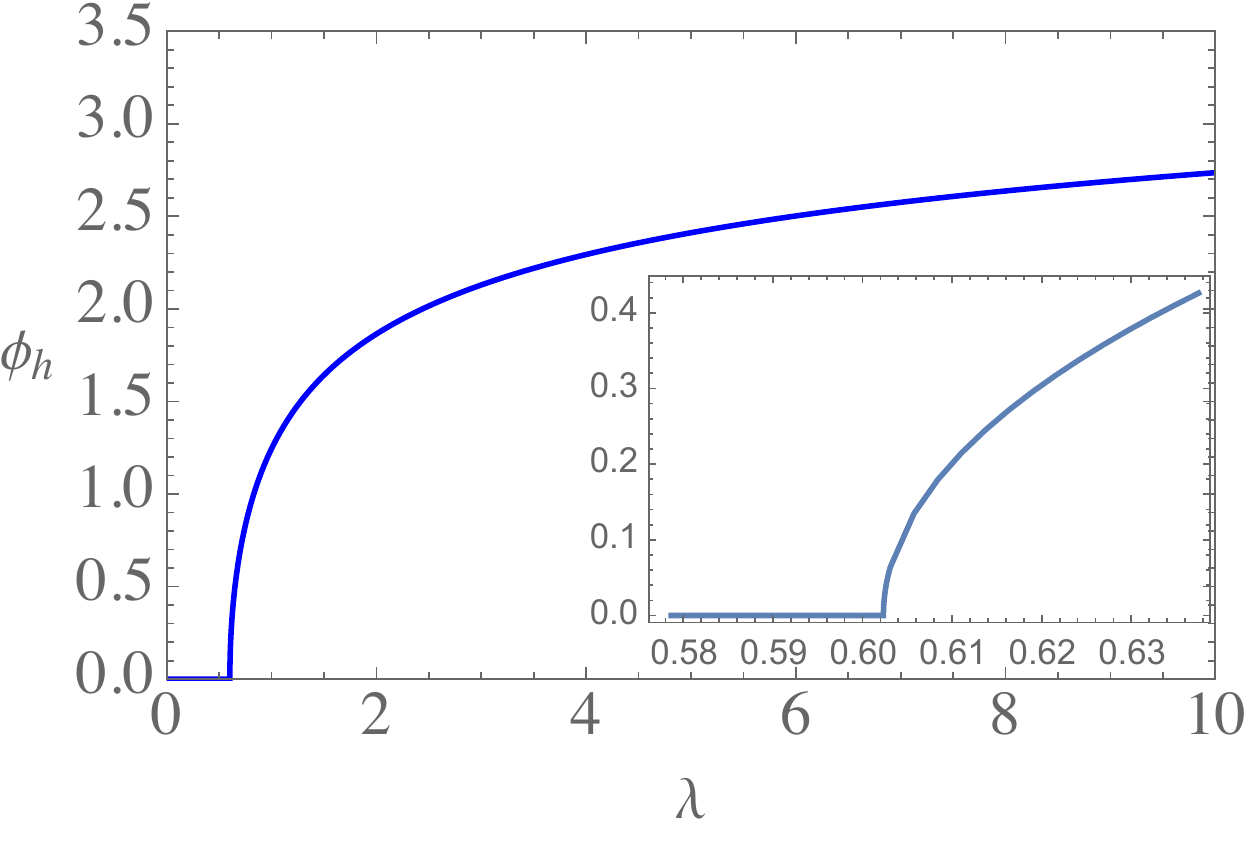}
  \includegraphics[scale=0.42]{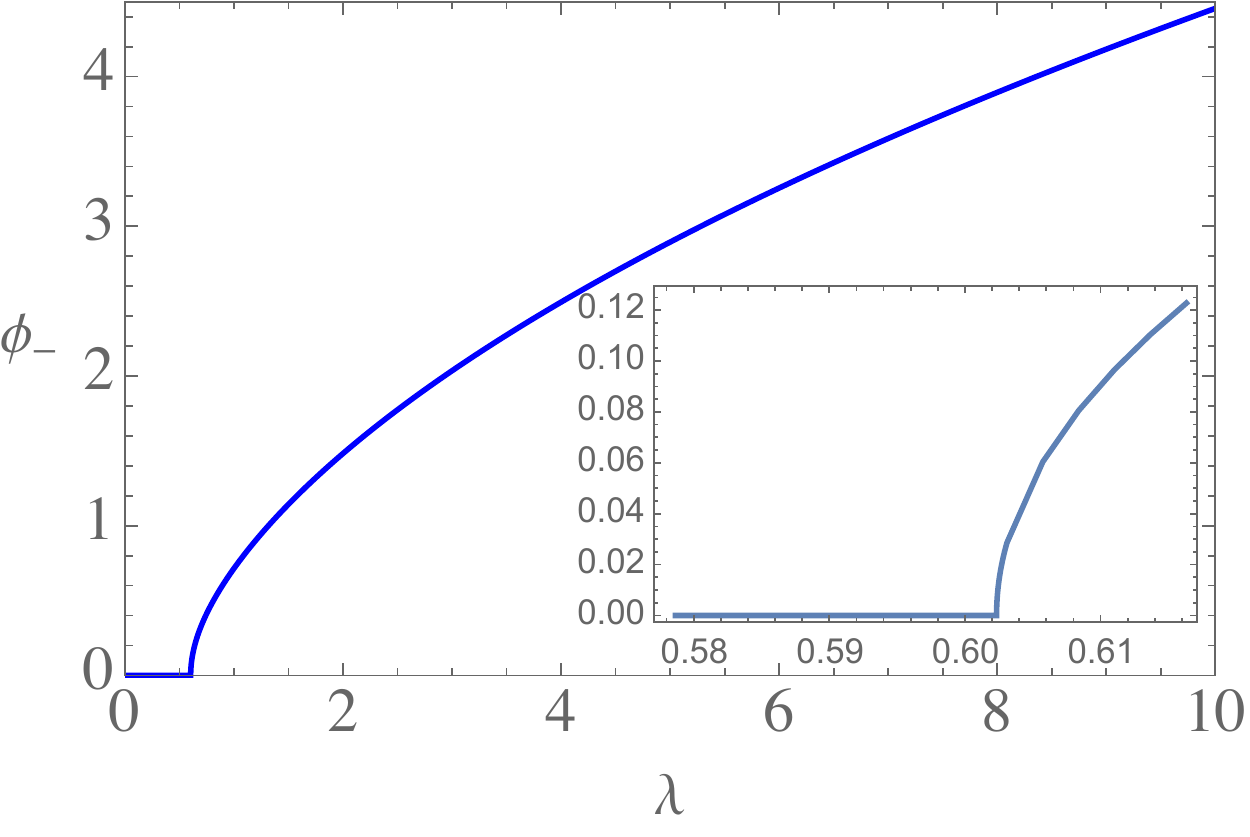}
  \caption{Left: the profiles of the scalar field in asymptotical flat  schwarzschild black hole with different values of $\lambda$. Middle: the value of scalar field near the event horizon $\phi_h$ as the function of $\lambda$. Right: the corresponding scalar charge of the black hole $\phi_{-}$ at infinity as the function of $\lambda$.}\label{schpro}}
\end{figure}

We then study the behaviour of scalar field by solving the perturbation equation \eqref{scalareq} and expect to see the signal of scalarization in the probe limit. By scanning the coupling parameter, it is found that there exists a threshold value $\lambda_t\simeq 0.60234$\footnote{Noted that previously we discussed that for $\lambda^2>\frac{5}{12}M^2$, i.e. $\lambda\gtrsim0.645$, the background black holes would be unstable. Here the threshold value $\lambda_t\simeq 0.60234$ is a bit smaller. This is reasonable because \eqref{condition} is only a sufficient condition, which was also verified in\cite{Doneva:2017bvd}.  }. When  $\lambda$ is smaller than the threshold value, the scalar field is zero everywhere which is trivial. When the coupling is larger than $\lambda_t$, the scalar hair would emerge and increase as the coupling increases. The profile of the scalar field with various values of $\lambda$ is shown in the left plot of Fig. \ref{schpro}. Moreover, in the middle plot of Fig. \ref{schpro}, we present the value of the scalar field near the event horizon $\phi_h$ as the function of $\lambda$ and corresponding scalar charge $\phi_{-}$ at infinity is shown in the right figure. The black line on the horizontal axis is the trivial scalar field. As the coupling parameter increases larger than $\lambda_t$, there emerges the bifurcation dividing the trivial solution, the non-trivial scalar field formes and grows dramatically.
This phenomenon indicates that due to the  instability brought by the GB coupling, the scalar field may emerge and grow  near the horizon of AFBH background, such that a new branch of scalarized hairy black hole develops. It is noticed that the scalarized hairy black hole solutions with backreaction were numerically explored in \cite{Doneva:2018rou}, and it was found that the entropy of the scalarized black hole is larger than of AFBH solution which indicates the hairy solution is physically favorable.

\subsection{Asymptotical AdS black hole}\label{ADBH}

In the AdS case with $\Lambda=-3/L^2<0$, the redshift function in the solution \eqref{metric} with spherical horizon is $g(r)=1-\frac{M}{r}+\frac{r^2}{L^2}$ ($M=1+\frac{1}{L^2}$ as $r_h=1$) and the GB term ${\cal R}^2_{GB}=\frac{4}{r^2}\left[g'(r)^2+(g(r)-1)g''(r)\right]$.

As is known, the AdS background could possess a classical instability for the real scalar field, but provided its effective mass is above the Breitenlohner-Freedman (BF) bound, the perturbation do not break the AdS behavior\cite{BFbound}. In our case, the effective mass  derived from \eqref{FE2} is given by $m_{eff}^2=m^2-\frac{\lambda^2}{2}{\cal R}^2_{GB}$, and so the regular tachyonic instability could exist as $m_{eff}^2>m_{BF}^2=-9/4L^2$.
Specially, near infinity, the Klein-Gordon equation becomes
\begin{equation}
	\phi''(r)+\frac{4}{r}\phi'(r)+\frac{-m^2L^2+\frac{12\lambda^2}{L^2}}{r^2}\phi(r)=0
\end{equation}
which gives the boundary behavior of the scalar field as
\begin{equation}\label{infinity}
	\phi(r\rightarrow\infty)=\phi_- r^{-\alpha_-}+\phi_+r^{-\alpha_+}.
\end{equation}
Here $\alpha_{\pm}=\frac{3\pm \sqrt{9+4m_e^2L^2}}{2}$ and $m_e^2L^2=m^2L^2-12\frac{\lambda^2}{L^2}$ is the asymptotical behavior of
$m_{eff}^2L^2$. It is obvious that $m_e^2>m_{BF}^2=-9/4L^2$ could also ensure the scalar field at infinity to be convergent to be zero. Thus, in AdS case, for the parameters satisfied the above analysis, we are safe to solve the probed perturbative equation \eqref{scalareq} to see how exactly the GB coupling and AdS curvature radius affect the fate of the scalar field.

For a proper comparison in the flat and AdS cases, we set $m_e^2L^2=-2$ and then the asymptotic behaviour becomes $\phi(r\rightarrow\infty)=\phi_-/r+\phi_+/r^2$, such that the scalar charge $\phi_-$ term has the same decay behavior as that in  flat case.
In the numeric, we set $\phi_+=0$ as the boundary condition. It is noted that in general, the asymptotic behaviour of the scalar charge term could be different in flat and AdS cases. Thus, to compare the signal of scalarization, we have to carefully control the variables and then we require the decay behavior of the scalar charge term is coincident. It means that we shall probe the scalar field with the same scalar charge but not the same scalar mass, which is reasonable because non-vanishing scalar charge implies the occurrence of scalarization and modifies the general relativity black hole\cite{Antoniou_2018}.
Actually, though the mass of scalar field $m^2L^2=m_e^2L^2-12\frac{\lambda^2}{L^2}$ in AdS case seems to be very different from $m=0$ in flat case, it is obvious that the mass in AdS case $m^2=-2/L^2+12\lambda^2/L^4$ go to zero as $L\to\infty$, which is the flat limit. Our numerical result will also support the analysis, We shall show soon that all properties for very large $L$ in AdS case recover those in flat case as expected.

\begin{figure}[thbp]
\center{
  \includegraphics[scale=0.42]{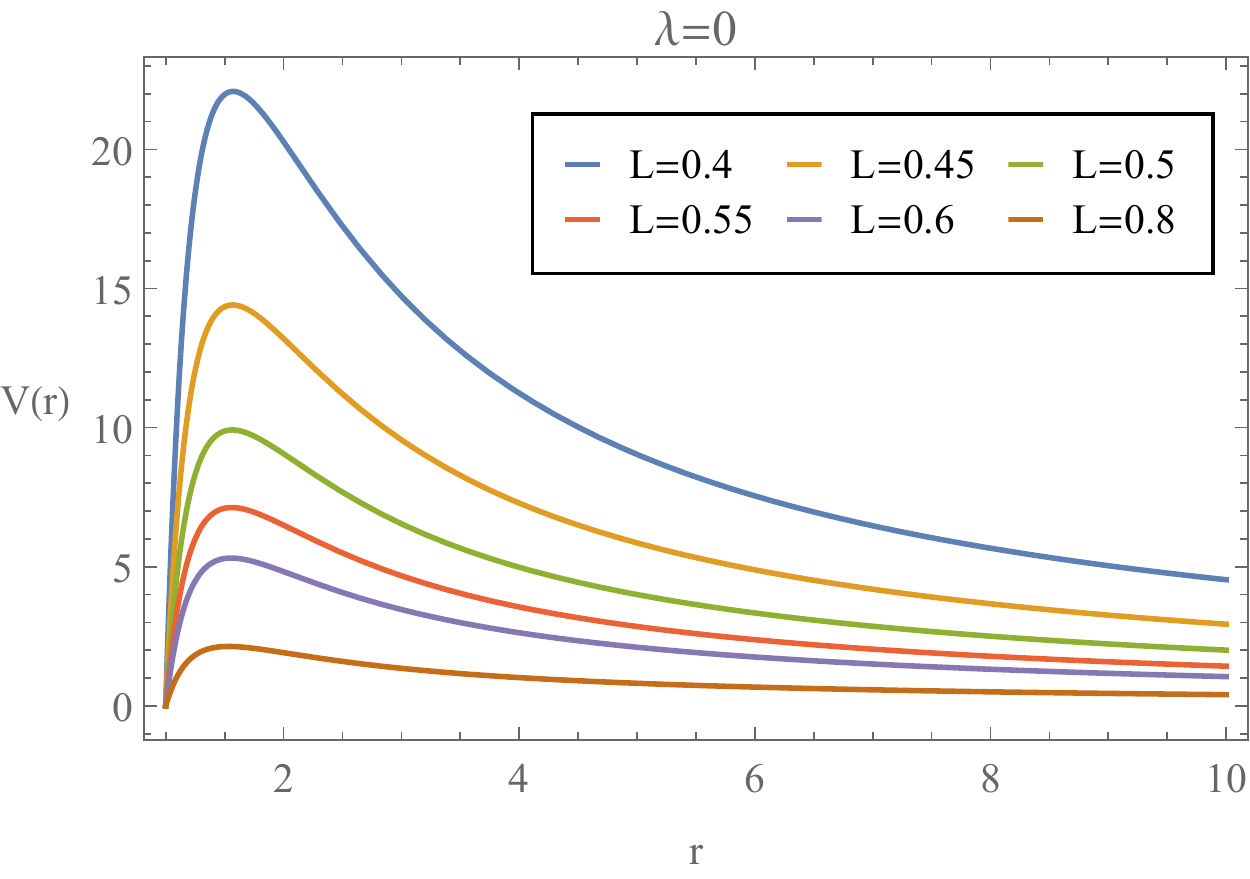}
  \includegraphics[scale=0.42]{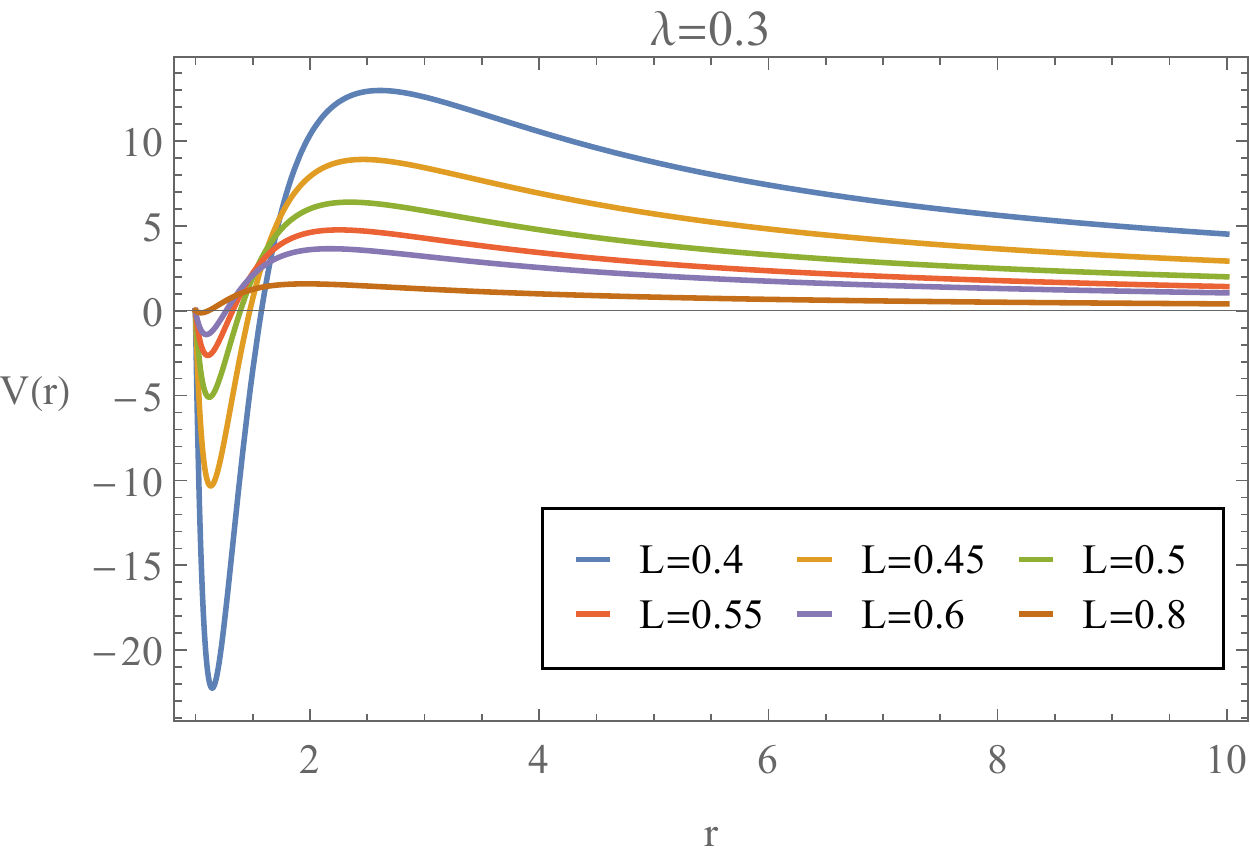}
  \includegraphics[scale=0.42]{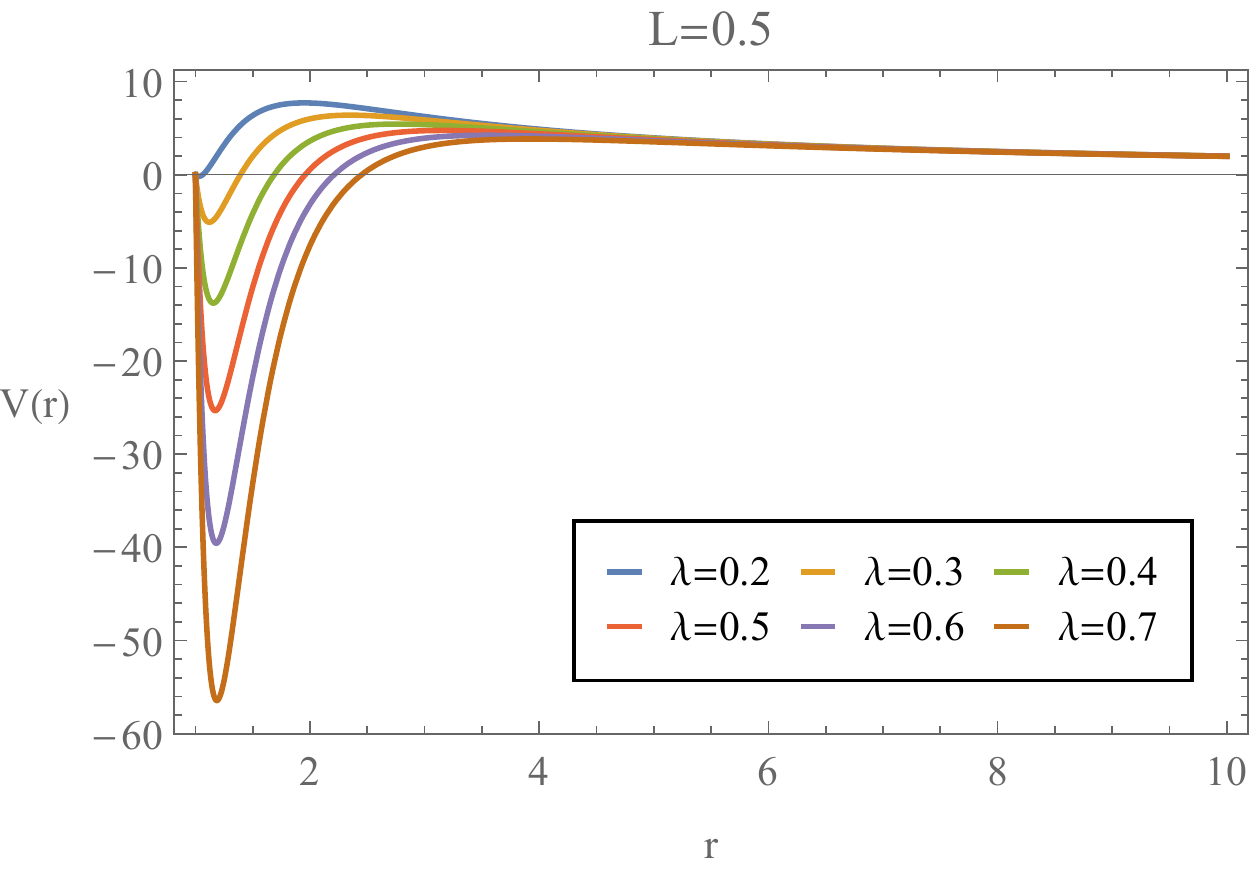}
  \caption{The effective potential in asymptotical AdS black hole with positive curvature.}\label{scapotenp1}}
\end{figure}

To proceed, we first show the potential function \eqref{eq-potential} in Fig. \ref{scapotenp1}. We see that the potential near the horizon could be negative for certain parameter regime and the negative potential may lead to  instability of the scalar field and a small scalar field perturbation may finally destabilizes  the AdS black hole background.
In the left plot without the GB coupling, the effective potential never becomes negative  unless the scalar field is charged  as addressed in \cite{Gubser1,Gubser2}. In the middle plot with $\lambda=0.3$, the potential well becomes deeper as $L$ decreases, meaning that there exists a critical $L_{*}$ that the potential is negative enough for the scalar field emerging near the horizon.  In the right plot with fixed $L=0.5$, as $\lambda$ increases, the potential becomes deeper. This behaviour is similar as that in AFBH case, however, by careful comparison, we find that for the same GB coupling, the well in the right plot is more deeper than that in AFBH case (see Fig. \ref{scapoten}). This implies that comparing to the flat spacetime, the scalar field in AdS spacetime would be easier bounded near the horizon, saying with smaller GB coupling.

\begin{figure}[thbp]
\center{
  \includegraphics[scale=0.42]{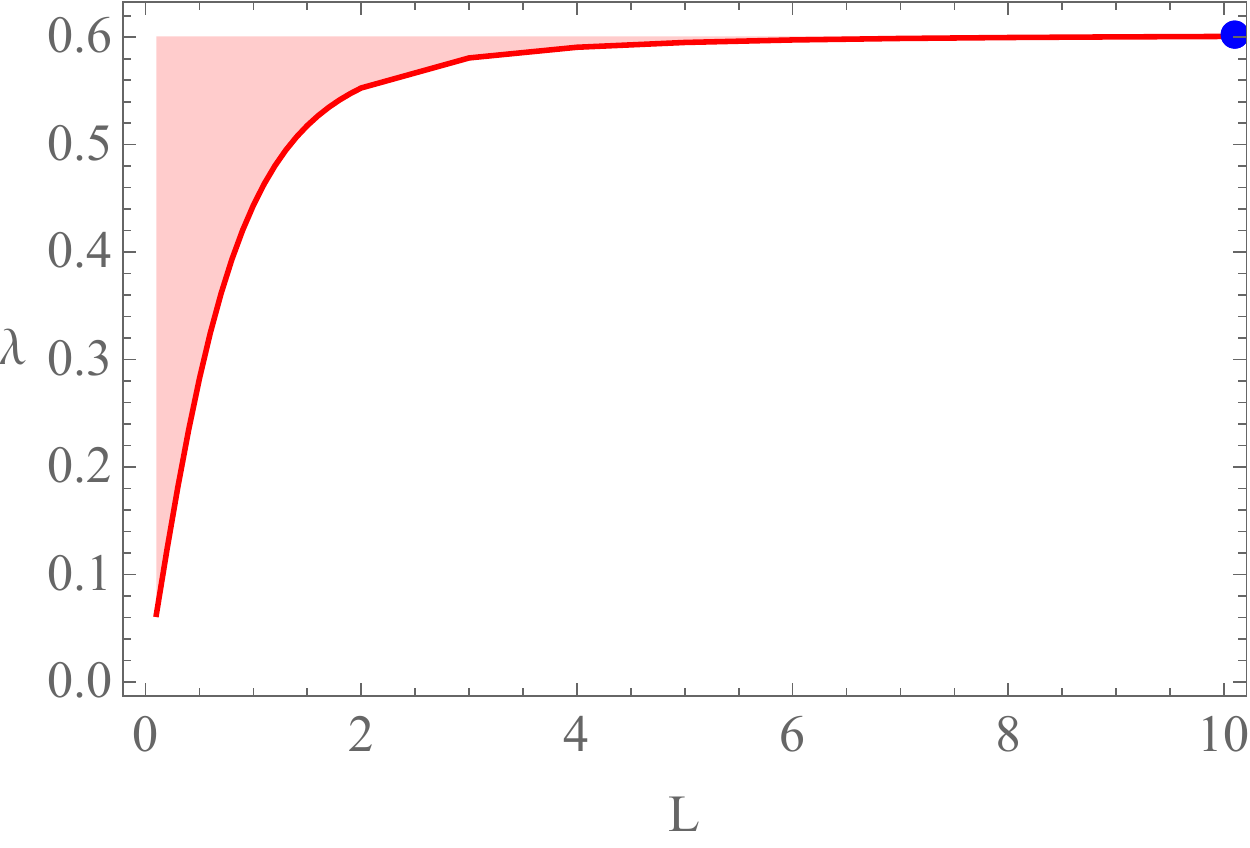}
  \includegraphics[scale=0.42]{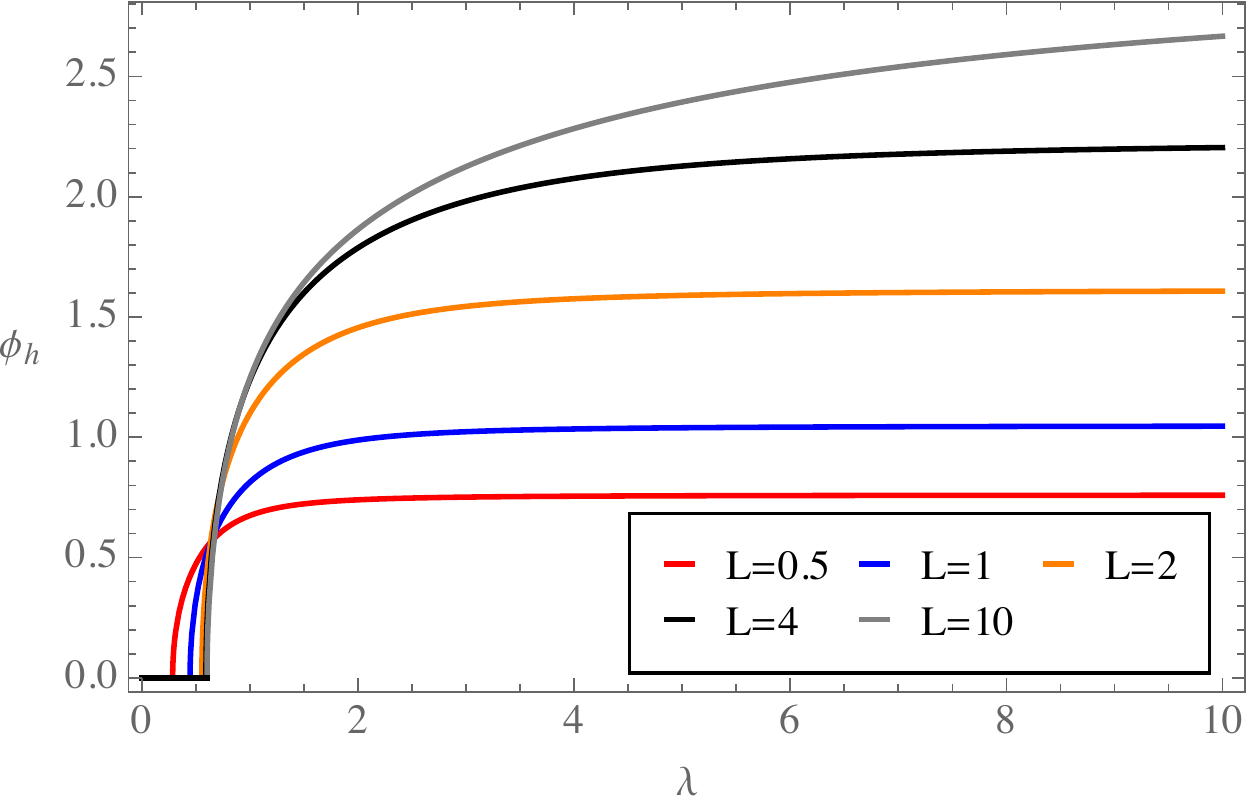}
  \includegraphics[scale=0.42]{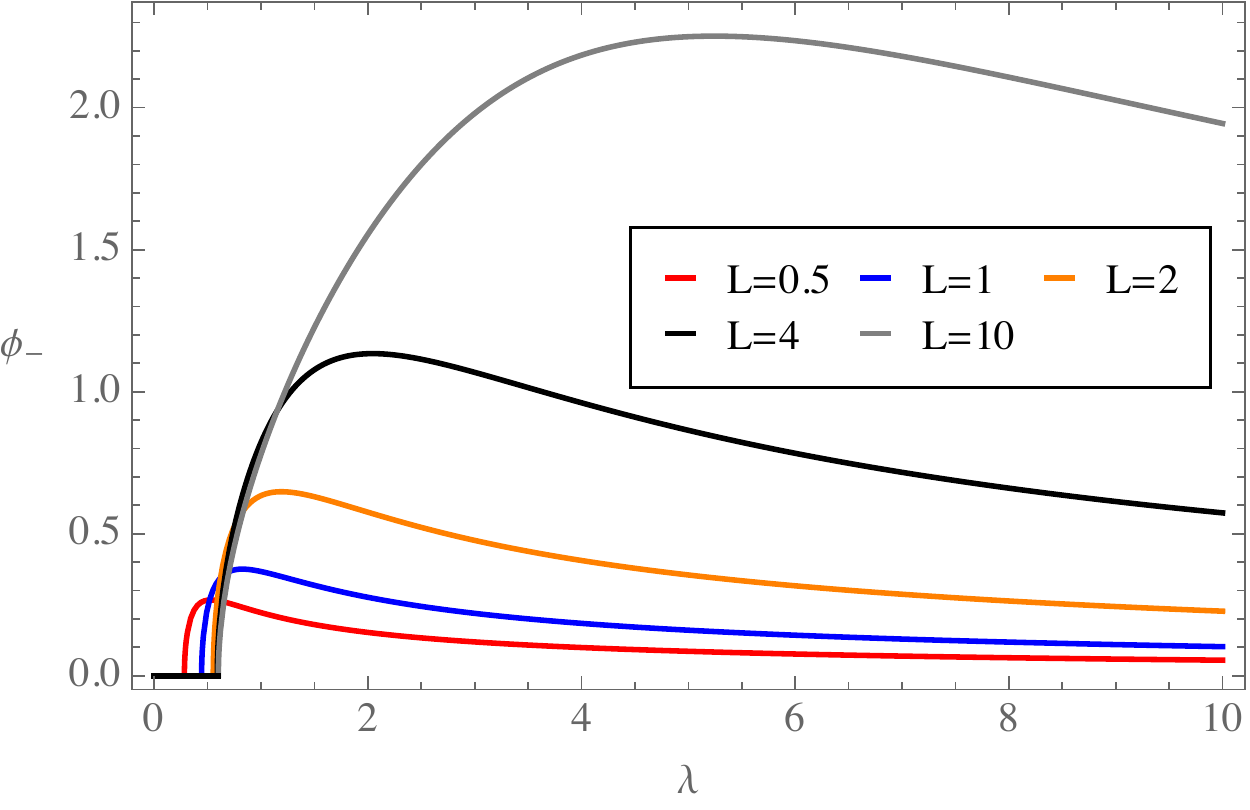}
  \caption{Left: the threshold curves for AdS black hole with positive curvature where the blue dot represents the threshold value of Schwarzschild black hole; Middle: the scalar hair of the black hole as the function of $\lambda$ with different AdS radius $L$; Right: the corresponding scalar charge of the black hole $\phi_{-}$ behaves with different AdS radius $L$.}\label{threscurve1}}
\end{figure}

Considering the scalar field near the horizon
\begin{equation}
\phi'(1)=\frac{L^2}{3+L^2}(m^2-6\frac{\lambda^2}{L^4}(3+2L^2+L^4)e^{-\phi(1)^2})\phi(1)~.
\end{equation}
and the boundary condition \eqref{infinity} by choosing $\phi_+=0$, we scan the parametric space $(\lambda,L)$ to find the solution of equation \eqref{scalareq}.
We find a critical line for the threshold of scalar field as shown on the left panel of Fig. \ref{threscurve1}. Below the critical line, the scalar field is zero which is trivial, while above the line the scalar field behaves as a non-trivial solution. We can extract two properties from the left panel of Fig. \ref{threscurve1}. Firstly, the threshold value of $\lambda$ first increases as $L$ becomes larger, and then tends to be a constant as $L\rightarrow\infty$. The constant is around  $\lambda_t\simeq 0.60234$ which matches the aforementioned value in AFBH case (see the blue dot). This is reasonable because as $L\rightarrow\infty$, the cosmological constant goes to zero which is the case we studied in the previous subsection.
Secondly, comparing to AFBH case, the threshold value of $\lambda$ in  AdS case is smaller, which implies that weaker GB coupling could introduce spontaneous scalarization for spherical AdS black hole. This behaviour supports our previous analysis on the effective potential of Fig. \ref{scapotenp1}.

In the middle panel of Fig. \ref{threscurve1}, we present the scalar field near the event horizon $\phi_h$ as a function of $\lambda$. For larger $L$, the value of $\lambda$ for the emergence of non-zero $\phi_h$ is larger, which is consistent with the observed on the left of Fig. \ref{threscurve1}. The scalar hair near the horizon emerges at the threshold coupling, and the value at the horizon increases as $\lambda$ increases and then tends to be constant. The scalar charge of the black hole $\phi_{-}$ shown in the right figure of Fig. \ref{threscurve1} supports our results. However, when $\lambda$ is small, $\phi_h$ decreases as $L$ increases, but when $\lambda$ is large enough, $\phi_h$ increases as $L$ increases.  Starting from the probe limit, we can obtain certain signal of spontaneous scalarization of the AdS background brought by GB coupling with the form $f(\phi)=\frac{1}{2}(1-e^{-\phi^2})$. Numerical methods are called for to handle the fully backreacted system and get the hairy solution, as was explored for various different $f(\phi)$ in \cite{Bakopoulos:2018nui}.


\subsection{Asymptotical dS black hole}

In the dS case with $\Lambda=3/L^2>0$, only the black hole with spherical horizon exists, so the redshift function in the solution \eqref{metric} is $g(r)=1-\frac{M}{r}-\frac{r^2}{L^2}$. As $r_h=1$, the mass is $M=1-\frac{1}{L^2}$, so that $L>1$ is required for the positive mass. Moreover, the cosmological horizon is located at $r_c=\frac{1}{2}(-1+\sqrt{4L^2-3})>r_h=1$, which further requires $L>\sqrt{3}$ in our setup.

Similar as the above subsections, we first discuss the validity of the probe approximation considering the scalar field as a test field in the dS spacetime. Following the analysis of \cite{Brihaye:2019gla}, we consider the Klein-Gordon equation Eq.(\ref{scalareq})  in pure dS background with $g(r)=1-\frac{r^2}{L^2}$ for which the cosmological horizon is located at $r_c=L$. By setting $\phi(r)=\frac{\varphi(r)}{r}$ and $x=\frac{r}{r_c}$, the Klein-Gordon equation is reduced as
\begin{equation}\label{A1}
	(1-x^2)\varphi''(x)-2x\varphi'(x)+u(u+1)\varphi(x)=0
\end{equation}
where $u=\frac{3\chi-1}{2}$, and
\begin{equation}
	\chi=\sqrt{1-\frac{4}{9}m^2r_c^2+4\frac{\lambda^2}{r_c^2}}\equiv
\sqrt{1-\frac{4}{9}\bar{m}^2_{eff}r_c^2}
\end{equation}
where we define $\bar{m}_{eff}^2=m^2-\frac{9\lambda^2}{r_c^4}$ in the second equality. We see that the Gauss-Bonnet coupling contributes to $\chi$ as a modification of that in \cite{Brihaye:2019gla}.

The solution of \eqref{A1} is
\begin{equation}\label{A3} \phi(\frac{r}{r_c})=\frac{\varphi(\frac{r}{r_c})}{r}=\frac{1}{r}P_u\big(\frac{r}{r_c}\big)+\frac{s}{r}Q_u\big(\frac{r}{r_c}\big)
\end{equation}
where $P_u$ and $Q_u$ are  Legendre functions and $s$ is an arbitrary constant. It is obvious that the solution diverges at $r=0$, and also  the second term diverges at $r=r_c$ because of $Q_u\big(\frac{r}{r_c}\big)$ diverges. So the regularity at $r=r_c$ requires $s=0$, then  the solution near $r_c$ is
\begin{equation}\label{A4}
\phi(r)\simeq \frac{1}{r_c}-\frac{\bar{m}_{eff}^2}{2}(r-r_c)+\mathrm{higher~ order ~term},
\end{equation}
while the solution at $r\gg r_c$ behaves as
\begin{equation}\label{A4}
	\phi(r)\simeq c_{+}r^{-\frac{3}{2}(1+\chi)}+c_{-}r^{-\frac{3}{2}(1-\chi)}, \ \ \ \ \mathrm{where} \ \ c_{\pm}=\frac{\Gamma(\mp\frac{3\chi}{2})}{\sqrt{\pi}\Gamma(1\mp\frac{3\chi}{2})}\big(\frac{r_c}{2}\big)^{\frac{1\pm 3\chi}{2}}.
\end{equation}
The precondition of scalarization is that a scalar field must have a tachyonic behaviour somewhere in between the BH and
cosmological horizon, which requires $\bar{m}^2_{eff}<0$ and $\chi>0$.
But this implies that the scalar field could diverge at $r\gg r_c$.
Therefore, the regularity of the scalar field at the cosmological horizon makes the tachyonic test field divergent at large distance, which breaks the probe approximation. Even though the probe limit may be valid between the event horizon and cosmological horizon, we can not check the signal of scalarization in probe limit with the same boundaries as working in flat and AdS cases. We could improve the numerical skills to involve the backreaction and do the comparison, and we hope to address this issue in the future.

\section{Signal of Scalarization of Black Holes with Negative Curvature}

\label{sec:k=-1}

In this section we will study the signal of scalarization if the background metric is a topological black hole (TBH).
For the action  $$ S=\int
d^{4}x\sqrt{-g}\Big{[} \frac{R+6L^{-2}}{16\pi G}\Big{]}~,$$
the presence of a negative cosmological constant allows the existence of black holes with a topology ${R}^{2} \times \Sigma$,
where $\Sigma$ is a two-dimensional manifold of constant curvature. The simplest solution of this kind,
when $\Sigma $ has negative constant curvature,  reads
\begin{equation}
ds^{2}=-\left( -1+\frac{r^{2}}{L^{2}}-\frac{2G\mu }{r}\right)
dt^{2}+\frac{ dr^{2}}{\left( -1+\frac{r^{2}}{L^{2}}-\frac{2G\mu
}{r}\right) }+r^{2}d\sigma^2_{2}\;,  \label{Top-BH-Einstein}
\end{equation}
where $d\sigma^2_{2}$ is the line element of $\Sigma $
which is just \eqref{metric} with $k=-1$, $\Lambda=-3L^{-2}$ and $M=2G\mu$.
The line element $d\sigma^2_{2}$ is locally
isomorphic to the hyperbolic manifold $H^{2}$ and $\Sigma $ must
be of the form
\[
\Sigma =H^{2}/\Gamma {\quad }\mathrm{{with\quad }\Gamma \subset
O(2,1)\;,}
\]
where $\Gamma $ is a freely acting discrete subgroup (i.e., without fixed points).

The configurations \eqref{Top-BH-Einstein} are asymptotically
local AdS spacetimes and these black holes are known as TBHs \cite{mann}-\cite{Cai:2001dz}. It has been  shown in
\cite{Gibbons:2002pq} that the massless configurations where
$\Sigma $ has negative constant curvature are stable under
gravitational perturbations and the  stability of the
topological black holes was discussed in \cite{Birmingham:2007yv,Wang:2001tk}. In a series of papers \cite{Koutsoumbas:2006xj,Koutsoumbas:2008pw,Koutsoumbas:2008yq} it was shown that there is evidence that a vacuum TBH goes over to a hairy configuration, the MTZ black hole \cite{Martinez:2004nb}, through a second-order phase transition. Therefore it is interesting to see if in the presence of the  coupling between scalar field and  the GB term, there is evidence of TBH  scalarization.

\begin{figure}[thbp]
\center{
  \includegraphics[scale=0.42]{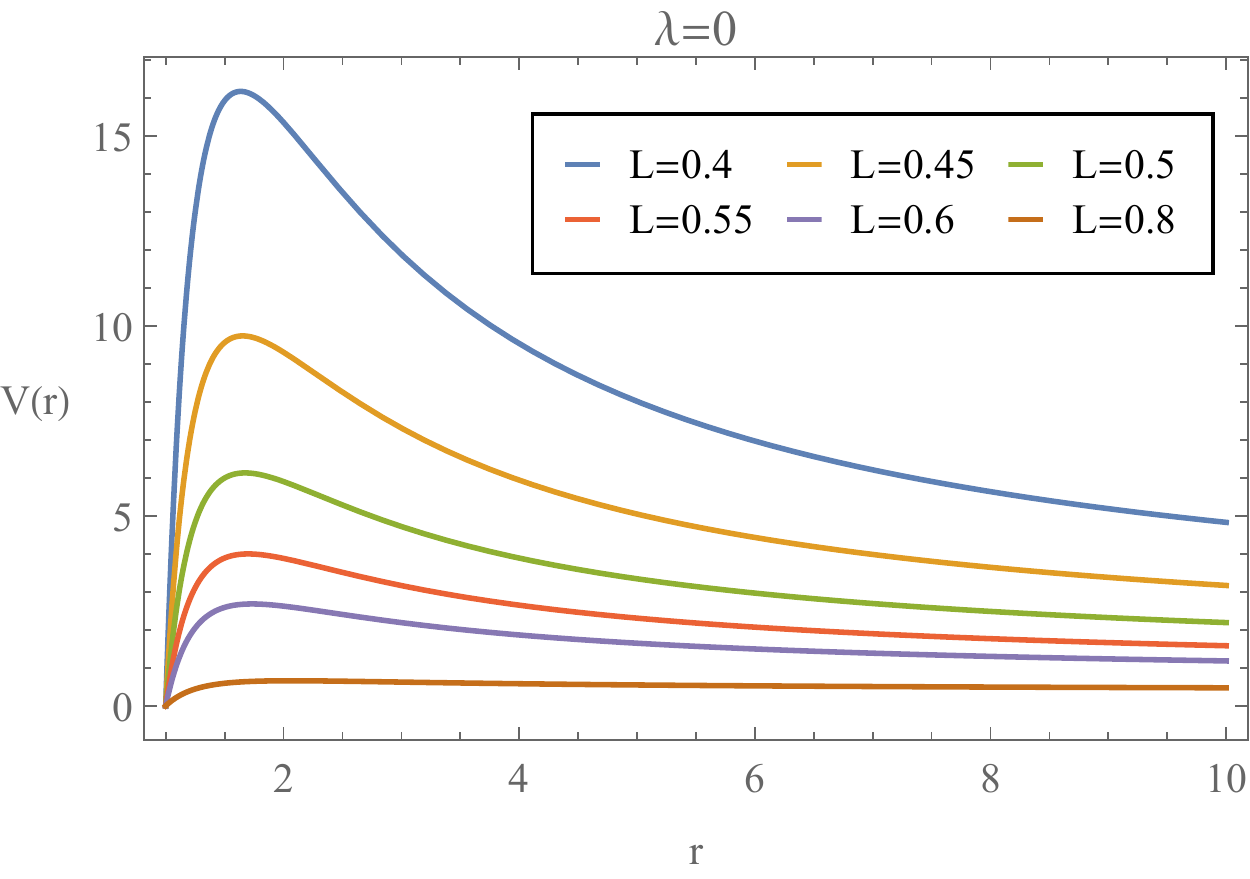}
  \includegraphics[scale=0.42]{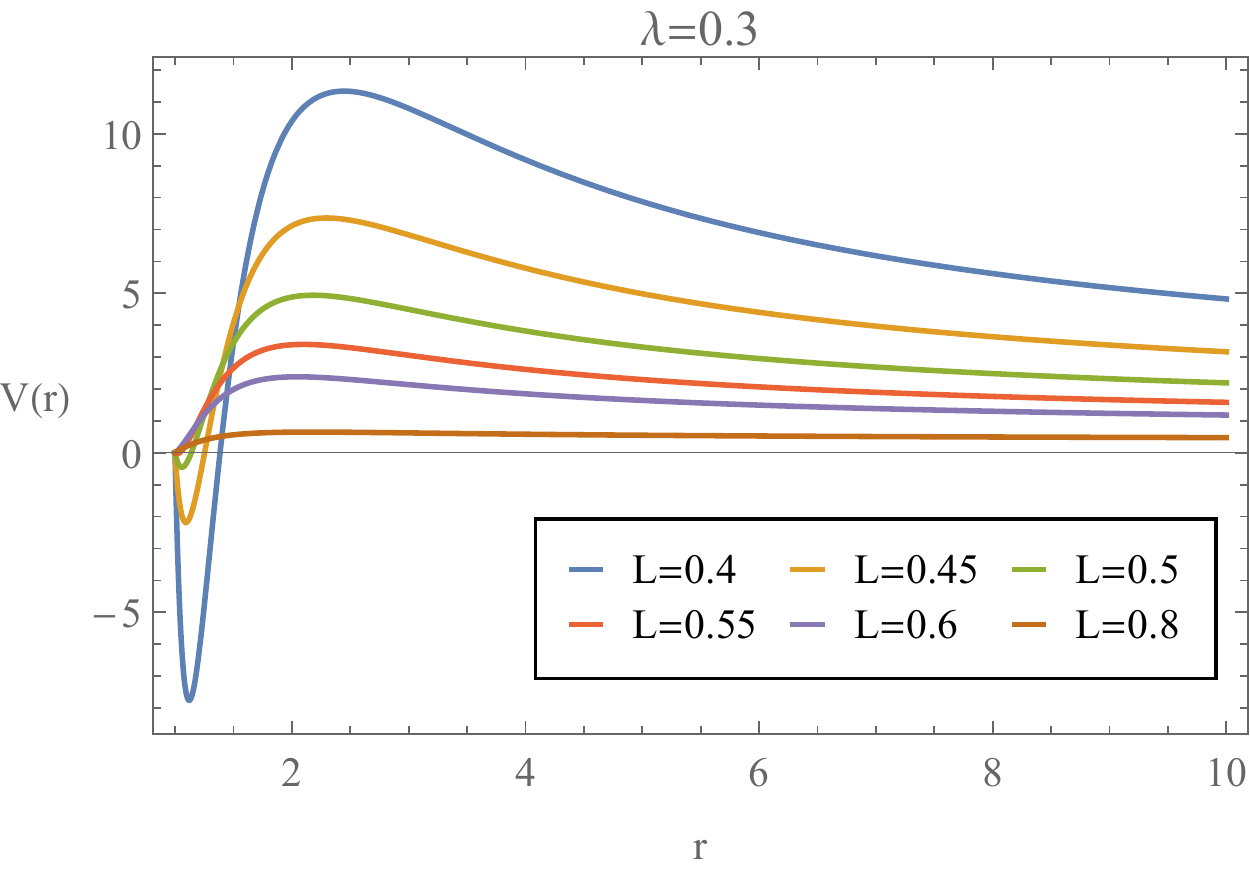}
  \includegraphics[scale=0.42]{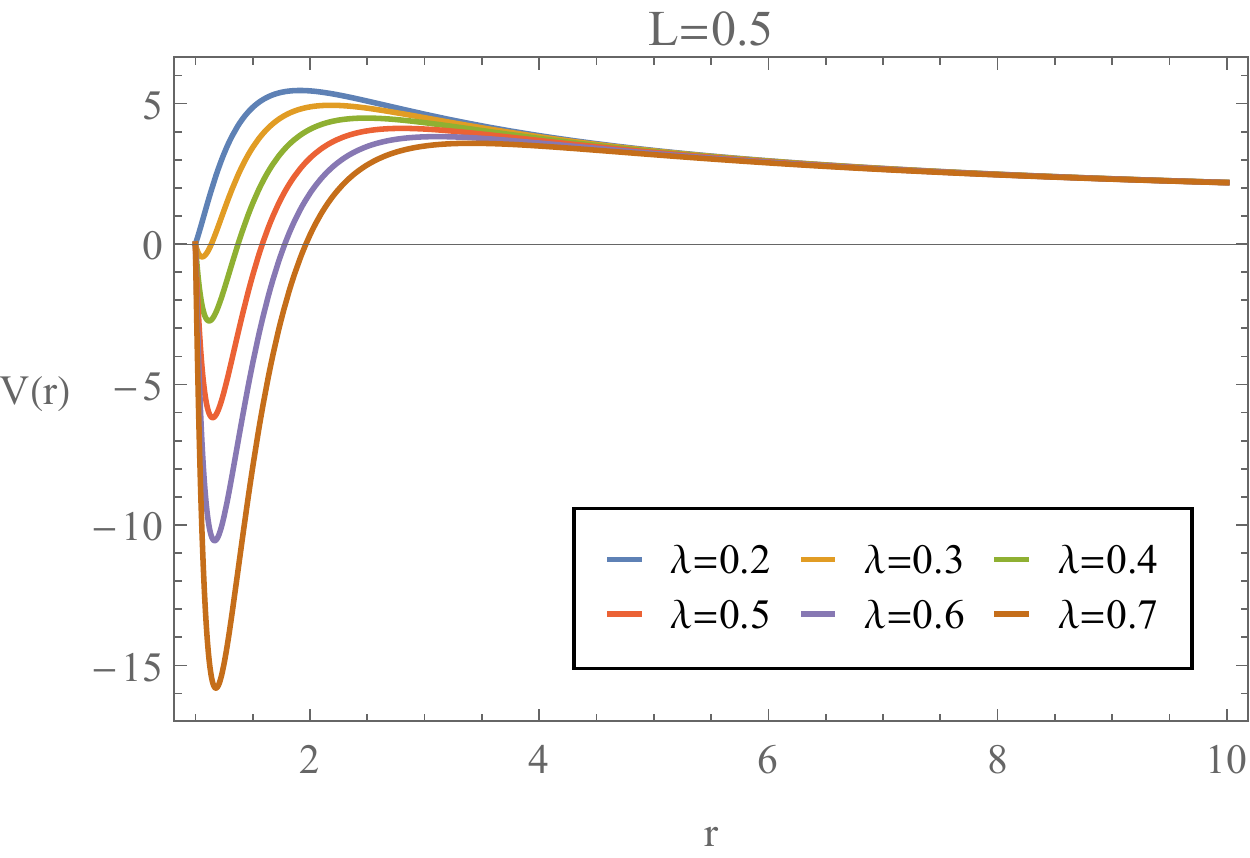}
  \caption{The effective potential for the AdS black hole with negative curvature.}\label{scapotenn1}}
\end{figure}

In this case, the GB term \eqref{GB} is evaluated as ${\cal R}^2_{GB}=\frac{4}{r^2}(g'(r)^2+(g(r)+1)g''(r))$.
To fulfill  $g(r_h)=0$, the black hole mass should satisfy  $M=r_h^3/L^2-r_h$. The positivity of black hole mass forces $0<L<r_h$, while the positivity of Hawking temperature requires $L<\sqrt{3} r_h$. Thus for $r_h\leq L<\sqrt{3} r_h$, the topological black hole have a positive Hawking temperature, but the black hole mass could be zero or negative, and the stability of the
these massless and negative mass black hole has been investigated in \cite{Birmingham:2007yv}. It is noted that the topological black holes have interesting structures in their own right, here for security we focus on topological black hole with positive mass, such that we have $0<L<1$ as we have set $r_h=1$.  Then the effective potential is shown in Fig. \ref{scapotenn1}.
Without the GB coupling (see the left plot), the effective potential is always positive, so there is no signal of tachyonic instability to induce the  scalarization, so that the TBH keeps stable. With $\lambda=0.3$ (see the middle plot), the potential well becomes deeper as $L$ decreases . When $L$ is small enough, the potential well would be deep enough to allow the tachyonic instability to happen. In the right plot with fixed $L=0.5$, as $\lambda$ increases, the potential well becomes deeper. We note that the rule for negative curvature in AdS case extracted from
Fig. \ref{scapotenn1} is qualitatively the same as that in the positive curvature case (see Fig. \ref{scapotenp1}). But with the same parameters, the potential well for $k=-1$ is shallower than that in the case with $k=1$. This means that the onset of spontaneous scalarization is easier to happen for AdS black hole with positive curvature than that with negative curvature.

Again, to further fix the parameters for the threshold of the nontrivial scalar field, we need to call back the perturbation equation \eqref{scalareq}.
Near the event horizon, the scalar field satisfies
\begin{eqnarray}
\phi'(1)=\frac{L^2}{3-L^2}(m^2-6\frac{\lambda^2}{L^4}(3-2L^2+L^4)e^{-\phi(1)^2})\phi(1)~.
\end{eqnarray}
While in asymptotical infinity, the behavior is also \eqref{infinity}. We shall also set $\phi_+=0$ in the numeric.

\begin{figure}[thbp]
\center{
  \includegraphics[scale=0.42]{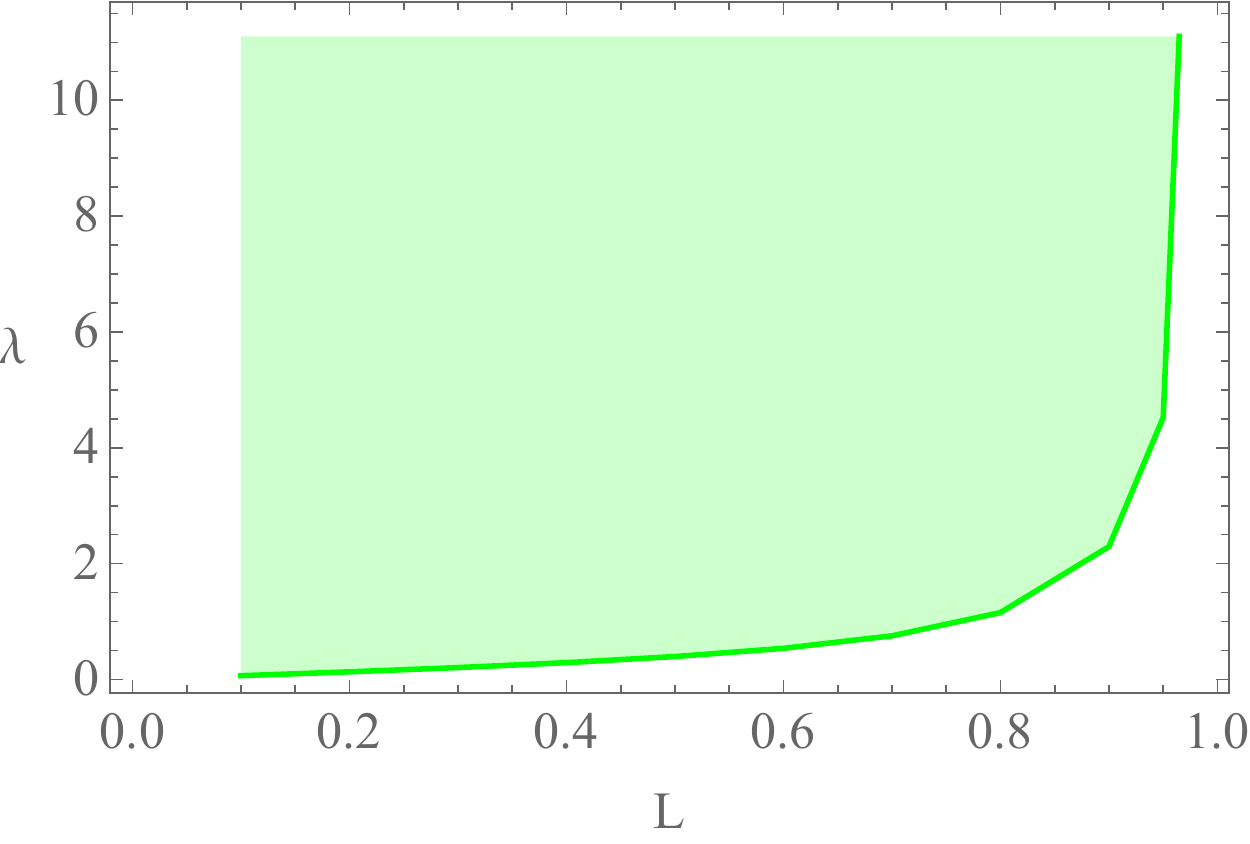}
  \includegraphics[scale=0.42]{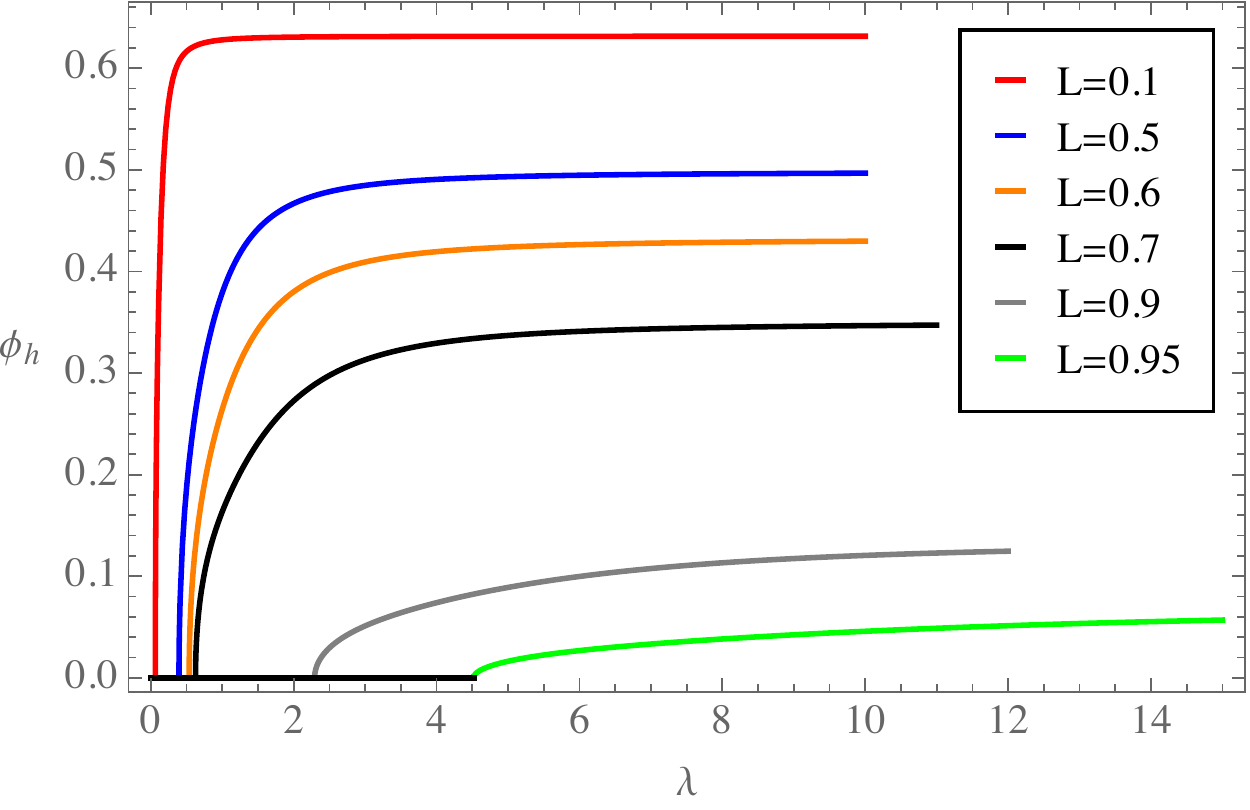}
  \includegraphics[scale=0.42]{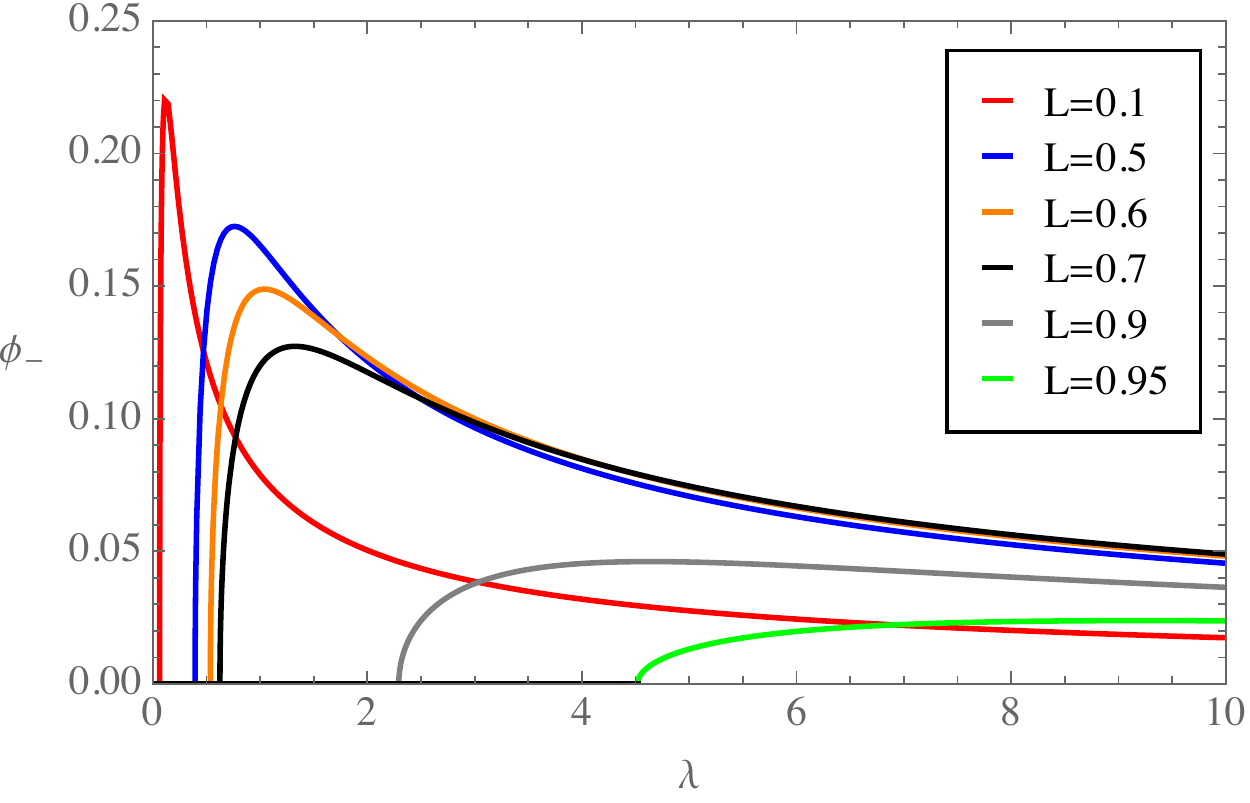}
  \caption{Left: the threshold curves  for AdS black hole with negative curvature; Middle: the scalar hair near the horizon as the function of $\Lambda$ with different AdS radius $L$; Right: the corresponding scalar charge of the black hole $\phi_{-}$ as the function of $\lambda$.}\label{threscurve-1}}
\end{figure}

The critical threshold parameters in $(\lambda, L)$ space is shown on the left of Fig. \ref{threscurve-1} where in the green shadow region the scalar field in nontrivial. As $L$ increases, the critical coupling also increases which is similar as that in the cases with $k=1$. The curve tends to be divergent as $L\rightarrow 1$ because there is no black hole solution when $L\geq 1$ in our setup. By comparing with left of Fig. \ref{threscurve-1} and Fig. \ref{threscurve1},  we find that the scalar hair is more difficult to form in the negative curvature than in the positive case.  This is consistent with the conclusion from the comparison of the potential function.

We also present the scalar hair near the horizon and the scalar charge at infinity as the function of $\lambda$ for different values of $L$  in the middle and right of Fig. \ref{threscurve-1}, respectively. With the increase of $L$, the threshold value $\lambda$ becomes larger, which matches the observation on the left of Fig. \ref{threscurve-1}. Moreover, as $L$ becomes larger, the curves become lower and lower rapidly and the  value of $\phi_h(\phi_-)$  decreases. Especially, when $L$ goes closely to the unity, $\phi_h$ becomes very small, which  indicates that the scalarization of the background black hole becomes very  difficult.
This behaviour is consistent with that in positive curvature at small coupling (see the middle plot of Fig. \ref{threscurve1}) as well as that in zero curvature shown in next section.

We have observed that scalarizations in  hyperbolic AdS black hole backgrounds depend on two factors, the coupling strength between the scalar field and the GB term and the cosmological constant through the radius $L$.  For the chosen $L$, once the coupling parameter $\lambda$ increases, the effective potential  in Fig. \ref{scapotenn1}  becomes deeper which can ignite tachyonic instability and start the hair formation more easily. When we have big enough $L$,  in  Fig. \ref{threscurve-1} we see that the required critical $\lambda$ to start the scalarization becomes bigger. In the middle panel of Fig. \ref{threscurve-1}, we find that for bigger $L$, the formed scalar hair is weaker even we have stronger coupling strength $\lambda$. The growth of $L$ counteracts the effect of the $\lambda$ and hinders the scalarization process.

\section{Signal of Scalarization of Black Holes with Zero Curvature}
\label{sec:k=0}

AdS black holes with  a flat space ($k=0$) is described as toroidal black hole spacetimes. Thus the corresponding redshift function in the metric \eqref{metric} is $g(r)=-\frac{M}{r}+\frac{r^2}{L^2}$ ($M=\frac{1}{L^2}$ as  $r_h=1$)
and the GB term
is ${\cal R}^2_{GB}=\frac{4}{r^2}(g'(r)^2+g(r)g''(r))$.

\begin{figure}[thbp]
\center{
  \includegraphics[scale=0.42]{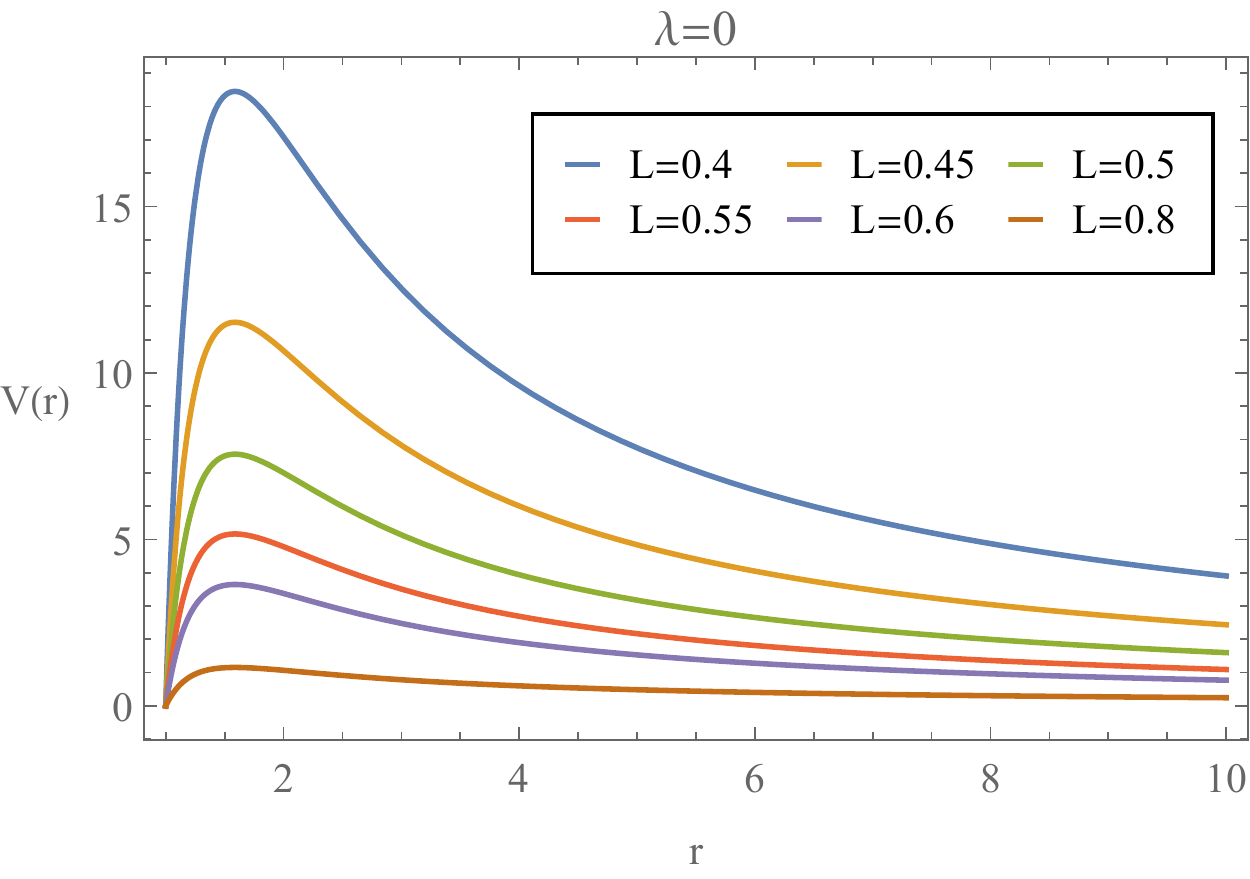}
  \includegraphics[scale=0.42]{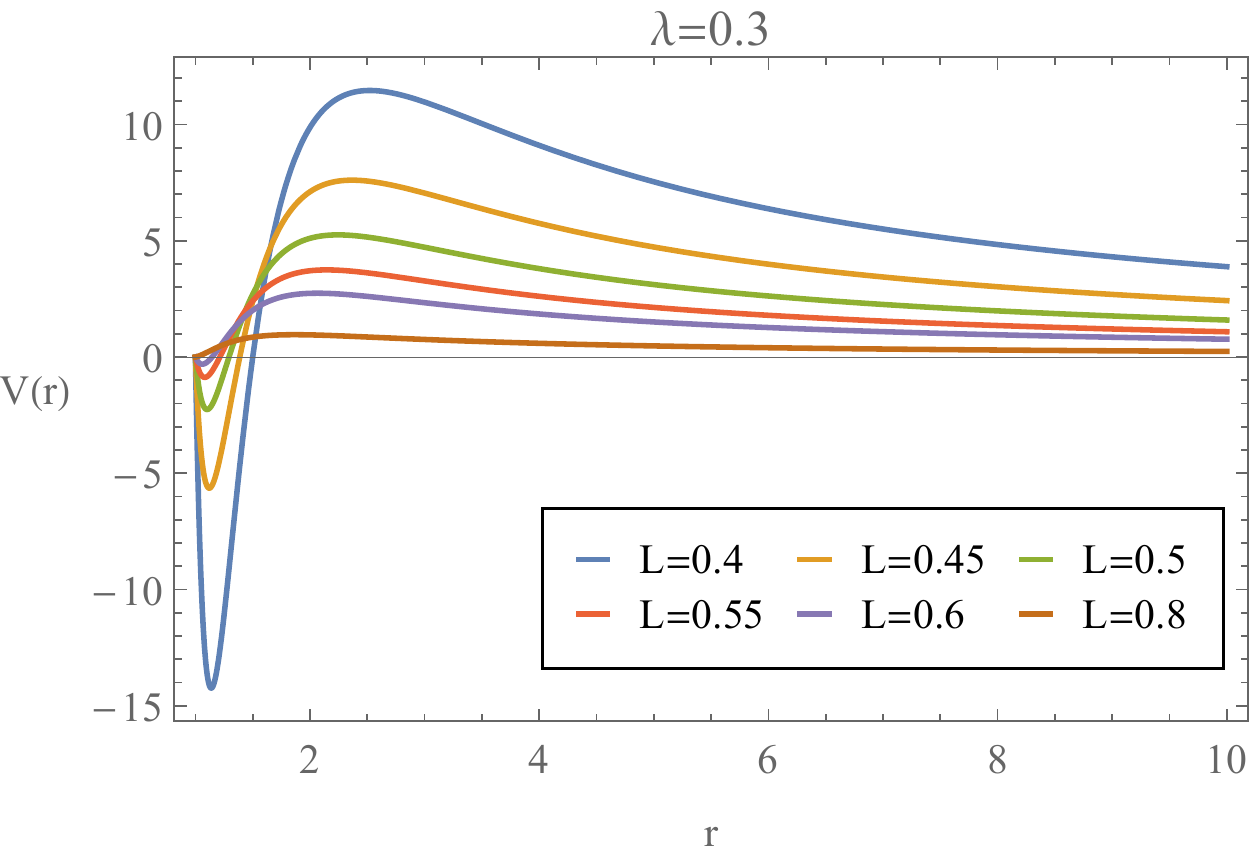}
  \includegraphics[scale=0.42]{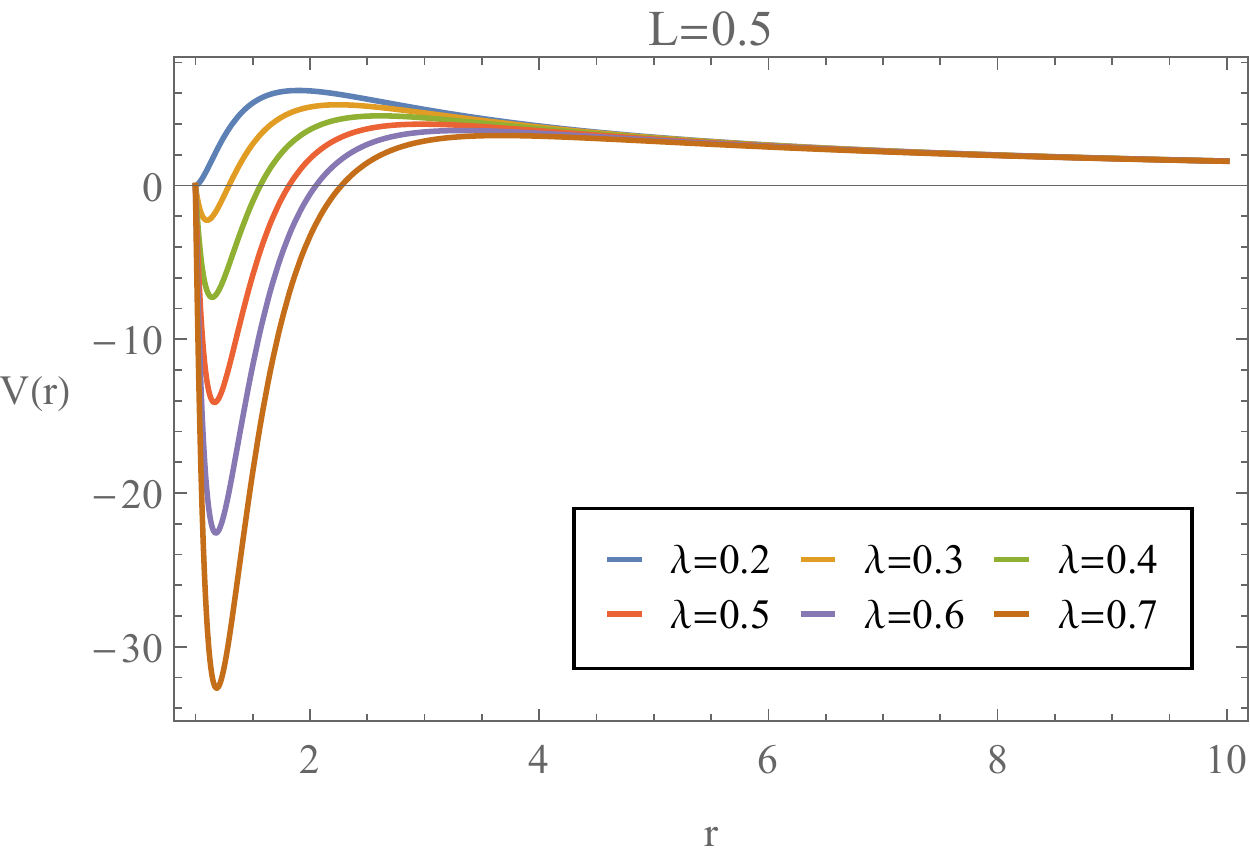}
  \caption{The effective potential for AdS black hole with zero curvature.}\label{scapotenz}}
\end{figure}

We plot the effective potential in Fig. \ref{scapotenz}. The potential function with $\lambda=0$  does not show negative well which insures the stability of the original toroidal hole. When the coupling between the scalar field and GB term is turned on,  for the chosen AdS radius $L$, the potential well near the horizon appears and becomes deeper as $\lambda$  increases. For the same coupling strength $\lambda$, similar potential well behavior will appear when we decrease the AdS radius $L$. The deep potential well can bound the scalar field near the horizon and lead to the instability of the original toroidal black hole.
This property is similar to those observed in the case with positive curvature  (Fig. \ref{scapotenp1}) and negative curvature (Fig. \ref{scapotenn1}). Through careful comparisons, we find that for the same parameters, the potential well in zero curvature case is deeper than that in the negative curvature case, while  shallower than that in the positive curvature background. This indicates that in ESGB theory, the horizon curvature influences the scalarization.  For AdS black hole with toroidal horizon the scalar hair can be formed more easily than the hyperbolic case, but if compared with the spherical AdS black hole, the scalarization process in toroidal AdS black hole is still more difficult.

To see the signal of the scalar field growing near the horizon, we solve the perturbation equation \eqref{scalareq} with the boundary condition near the horizon
\begin{eqnarray}
 \phi'(1)=\frac{L^2}{3}(m^2-18\frac{\lambda^2}{L^4}e^{-\phi(1)^2})\phi(1)~,
\end{eqnarray}
The behavior is also \eqref{infinity} in asymptotical infinity, and we set  $\phi_+=0$ as usual in the numeric.

\begin{figure}[thbp]
\center{
  \includegraphics[scale=0.42]{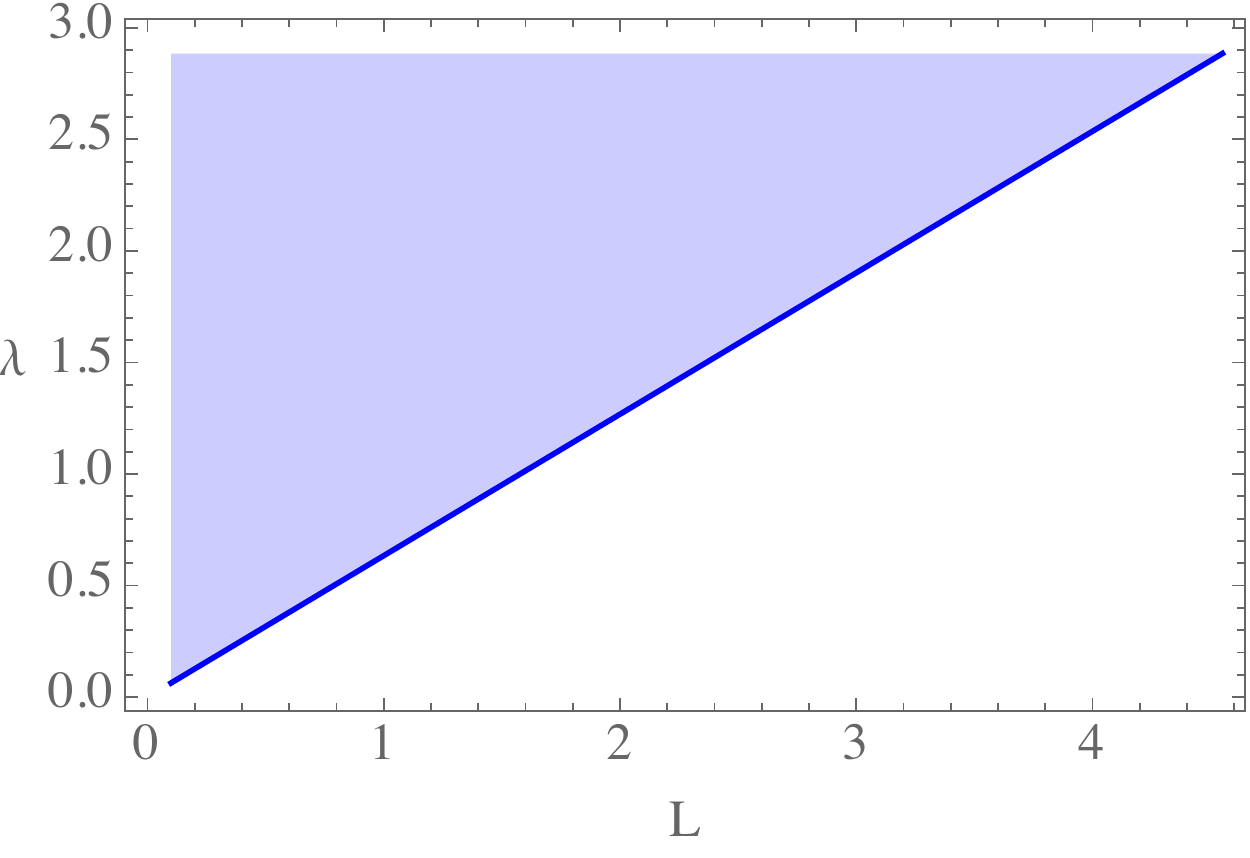}
  \includegraphics[scale=0.42]{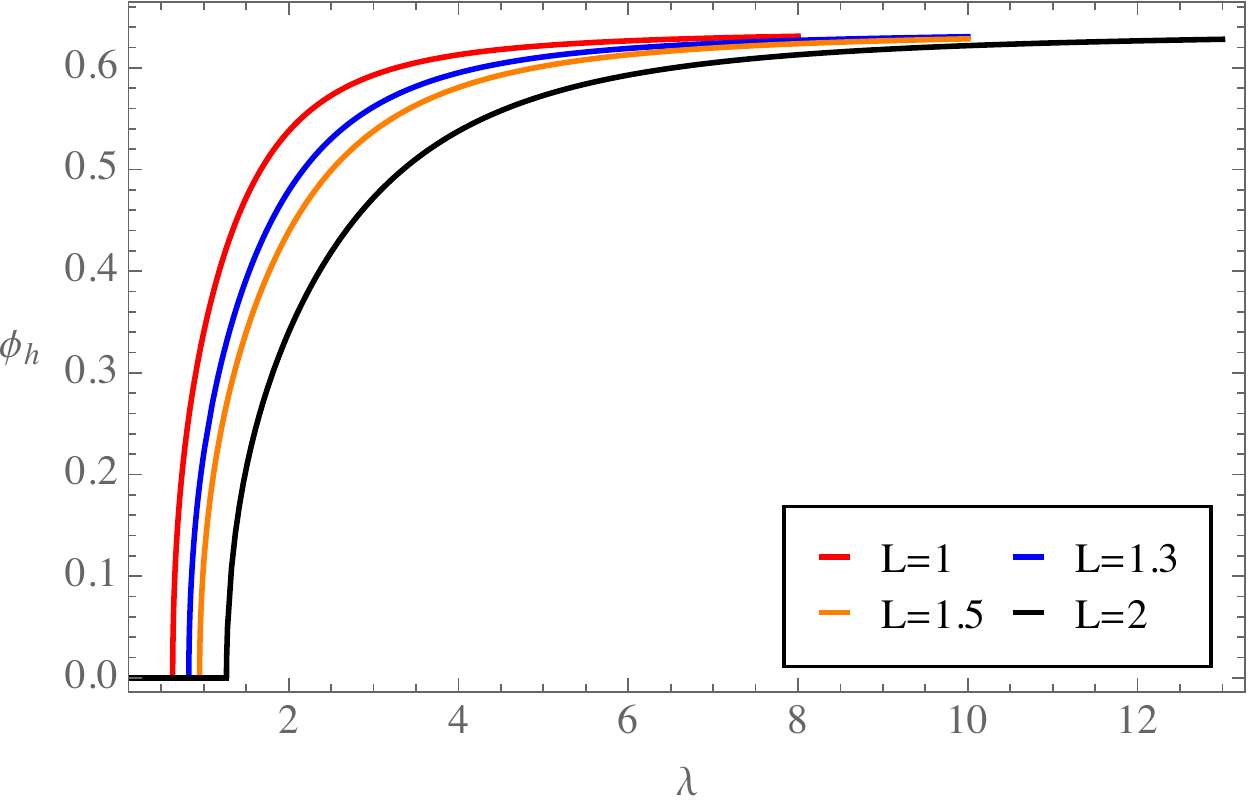}
  \includegraphics[scale=0.42]{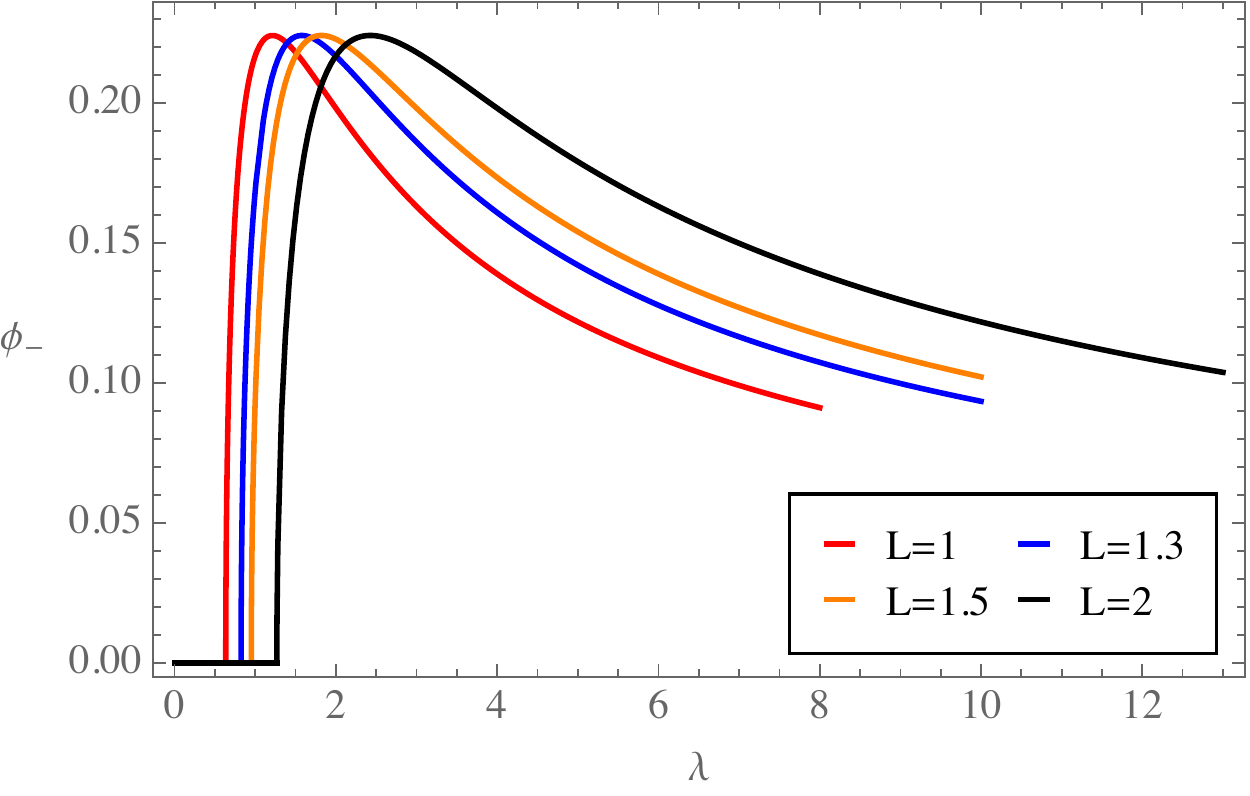}
  \caption{Left: the threshold curves for AdS  black hole with zero curvature; Middle: the scalar hair near the horizon as the function of $\lambda$ with different AdS radius $L$; Right: the corresponding scalar charge of the black hole $\phi_{-}$ as the function of $\lambda$.}\label{threscurve0}}
\end{figure}

The phase diagram in parameters $(L,\lambda)$ space is shown in the left panel of Fig. \ref{threscurve0} where below the blue line, the scalar field perturbation will trivially die out, but above this line the black hole scalarization will happen.  The values on this line are critical  threshold values to ignite the scalarization. When the AdS radius $L\rightarrow\infty$, we see that the threshold coupling $\lambda$ linearly approaches infinity, which means that the toroidal AdS black hole cannot be scalarized in such limits. In the limit $L\rightarrow\infty$, the background returns to the flat spacetime and there is no black hole with the zero curvature.

With some fixed $L$, we study the scalar field near the horizon and the scalar charge at infinity as the function of $\lambda$, and the results are shown in the middle and right of
Fig. \ref{threscurve0}, respectively.
Note that for large GB coupling, the probe limit may be not reasonable and backreaction should be involved.
But in some sense our result for small GB coupling is reliable and we see the signal of scalarization from the scalar field perturbation. In \cite{Guo:2020sdu}, we constructed the scalarized hairy black hole with planar horizon in AdS space by considering the full backreaction. There we have set $L=1$ and the threshold value is $\lambda\simeq0.64$ which is consistent with that in Fig. \ref{threscurve0} via careful comparison.

\section{Closing remarks}
In this paper, we discussed the signal of scalarization for black holes with different curvature topologies in various spacetimes. We considered  the scalar field as a probe to the general relativity solution in Einstein-scalar-Gauss-Bonnet theory. Our discussion is based on the analysis of its effective potential and the onset of nontrivial solution to the perturbed scalar equation.

We studied the signal of black hole scalarization with positive curvature in asymptotical flat, AdS and dS spacetime. We first  analyzed the validity of the probe approximation in various spacetimes. We found that in flat and AdS cases, a regular condition of test scalar field near event horizon could correspond to convergent (vanishing) condition at the asymptotical infinity, meaning the test field approximation could be valid; while in dS case, the regular condition near cosmological horizon leads to divergence at large distance, so that the test field approximation breaks down.
We then compared the effect of asymptotical flat and AdS spacetime spacetime structures on the scalarization process. Without the GB coupling $\lambda$, the effective potential of the scalar field are alway positive and no well emerges. As we increased the coupling, the negative potential well would show and becomes deeper which may lead to tachyonic instability. With the same parameters, the potential well for AdS black hole is deeper than that for Schwarzschild black hole, meaning that the scalar hair around the spherical AdS black hole may be easier to form.  We solved the perturbed scalar field equation in the parameter space
$(\lambda, L)$, and obtained the critical line above which the non trivial scalar hair can be found. Fig. \ref{threscurve1} shows that the scalar cloud around AdS black hole could form easier than that around the Schwarzschild black hole. Moreover, in the limit $L\rightarrow\infty$, the threshold coupling in AdS case approaches the value in Schwarzschild black hole from below as shown in the left of Fig. \ref{threscurve1}.


We also investigated the signal of AdS black hole scalarization process with different curvatures of the horizon.
For AdS black hole with positive $(k=1)$, negative $(k=-1)$ and zero $(k=0)$  curvatures, the properties of effective potential is qualitatively the same (see Fig. \ref{scapotenp1}, Fig. \ref{scapotenn1} and Fig. \ref{scapotenz}, respectively), that's to say, no negative potential well appears with $\lambda=0$ for instability. The potential well near horizon appears and becomes deep as $\lambda$ (or $L$) increases (decreases), which could bound the scalar field near the horizon and lead to instability. However, the potential well for zero curvature is deeper than that for negative curvature but shallower than that for positive curvature. Notice that the effect of horizon curvature on the potential could be roughly analytically evaluated from the expression \eqref{eq-potential}, i.e, $V(r)=g(r)\left(\frac{g'(r)}{r}+\frac{A}{r^2}+m^2-\frac{\lambda^2}{2}{\cal R}^2_{GB}\right)$. With fixed $L$ and $\lambda$, we can diagnose the potential as follows. (i) $m^2=-2/L^2+12\lambda^2/L^4$ for different horizon curvatures is the same in our setup. (ii) For the lowest mode, the topological term $A$ satisfies $A(k=1)=A(k=0)=0<A(k=-1)=1/4$ . (iii) Fixing horizon radius implies that the black hole mass satisfies $M(k=1)>M(k=0)>M(k=-1)$, then it is easy to obtain that the group term $T=\frac{g'(r)}{r}-\frac{\lambda^2}{2}{\cal} R^2_{GB}$ in the potential satisfies $T(k=1)<T(k=0)<T(k=-1)<0$ for fixed $r$ near horizon. (iv) The overall factor $g(r)$ fulfills $g(k=1)>g(k=0)>g(k=-1)>0$ for fixed $r$ outside the horizon. Combining (i)-(iv), it is not difficult to reduce that the effect of horizon curvature on the negative potential well  is $V(k=1)<V(k=0)<V(k=-1)$ near horizon.  It is noted that only the angular eigenvalues term $A$ mentioned in point (ii) is truly the topological effect, and the terms mentioned in points (iii-iv) are actually geometrical  effect. Thus, in this sense, the effect from the horizon curvature on the potential as well as the scalarization is reflected by the interplay between the topology and geometry.

\begin{figure}[thbp]
\center{
  \includegraphics[scale=0.6]{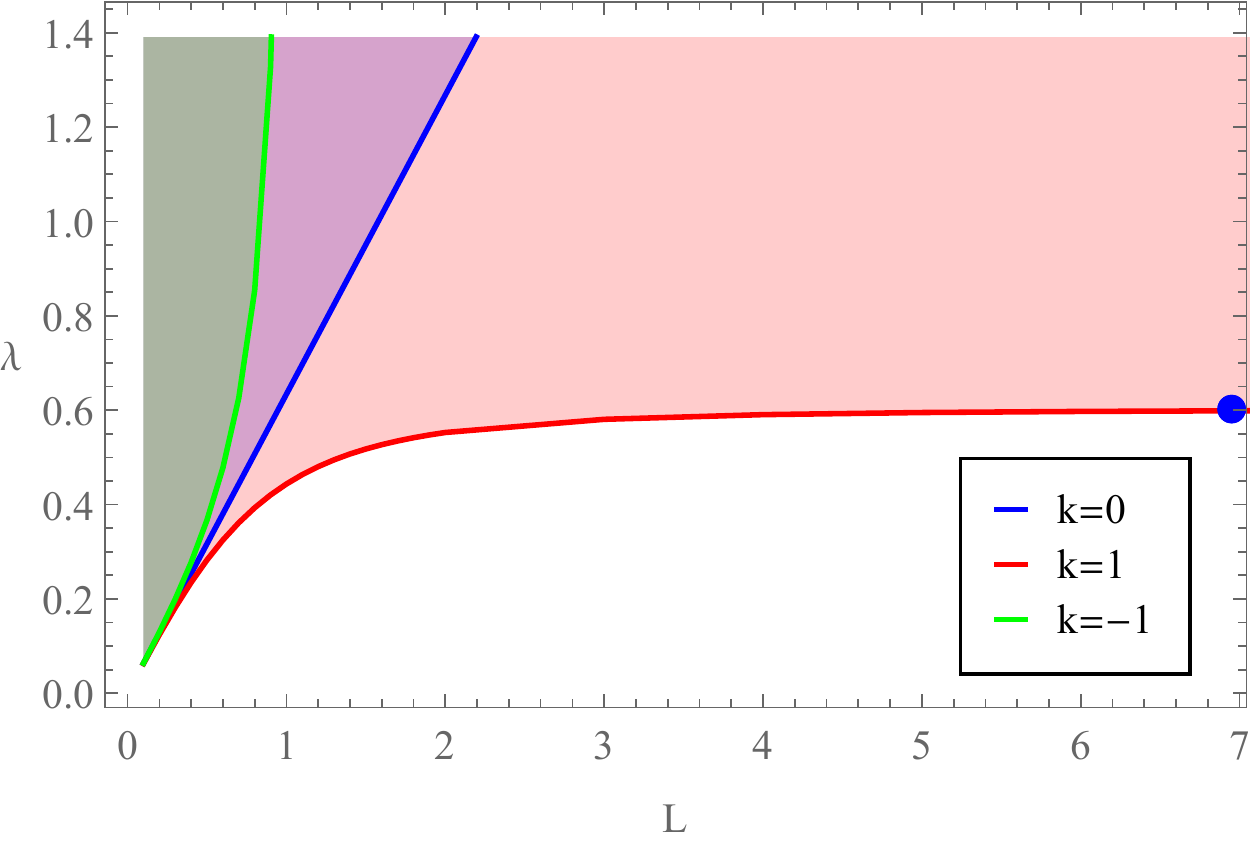}
  \caption{The threshold curves for AdS black hole with different curvatures. The blue dot represents the threshold value of Schwarzschild black hole.}\label{threscurve}}
\end{figure}

The behavior of potential indicates that in ESGB theory, the scalar hair around AdS black hole with toroidal horizon may be easier to be formed than that with hyperbolic horizon  but more difficult than that with  spherical horizon.
Moreover, we collected the critical threshold curves for different curvatures in Fig. \ref{threscurve} where the blue dot represents the threshold value of Schwarzschild black hole in the limit $L\rightarrow\infty$. It is obvious that for the AdS black hole with fixed $L$, the scalar hair is the easiest to be formed around spherical horizon, then around the toroidal horizon and it is the most difficult to be bounded near hyperbolic horizon.

We shall present two comments on Fig. \ref{threscurve}. $i)$ In the limit $L\rightarrow 0$, the threshold coupling in all three cases is very small in which case the scalarization could easily realized. Moreover, we know from the effective potential that the GB coupling should be involved  for the tachyonic instability and it further leads to the spontaneous scalarization. So we may argue that the threshold value would be tiny as $L\rightarrow 0$, or the scalarization could cut off at certain tiny  $L$. This direction with large $\Lambda$ deserves further study.
$ii)$ In the limit $L\rightarrow\infty$, the threshold coupling for zero or negative curvature are divergent, suggesting that the scalar field can not be bounded in these cases. This is understandable because as $L\rightarrow\infty$, the spacetime become flat and only
Schwarzschild black hole with positive curvature exists as background in probe limit. Moreover, it is interesting to involve the full backreaction of the scalar field into the geometry and verify our results.


\begin{acknowledgments}
This work is supported by the Natural Science Foundation
of China under grant Nos.11705161, 12075202 and 11690022. Xiao-Mei Kuang is also supported by Fok Ying Tung Education Foundation under grant No.171006.
\end{acknowledgments}


\begin{thebibliography}{99}
\baselineskip=0.5 cm

\bibitem{GW1}
  B.~P.~Abbott {\it et al.} [LIGO Scientific and Virgo Collaborations],
  ``Observation of Gravitational Waves from a Binary Black Hole Merger,''
  Phys.\ Rev.\ Lett.\  {\bf 116}, no. 6, 061102 (2016)
  [arXiv:1602.03837 [gr-qc]].

\bibitem{GW2}
  B.~P.~Abbott {\it et al.} [LIGO Scientific and Virgo Collaborations],
  ``GWTC-1: A Gravitational-Wave Transient Catalog of Compact Binary Mergers Observed by LIGO and Virgo during the First and Second Observing Runs,''
  Phys.\ Rev.\ X {\bf 9}, no. 3, 031040 (2019)
  [arXiv:1811.12907 [astro-ph.HE]].

\bibitem{GW3}
  B.~P.~Abbott \textit{et al.} [LIGO Scientific and Virgo],
  ``GW190425: Observation of a Compact Binary Coalescence with Total Mass $\sim 3.4 M_{\odot}$,''
  Astrophys. J. Lett. \textbf{892} (2020) no.1, L3
  [arXiv:2001.01761 [astro-ph.HE]].

\bibitem{EHT}
  K.~Akiyama \textit{et al.} [Event Horizon Telescope],
  ``First M87 Event Horizon Telescope Results. I. The Shadow of the Supermassive Black Hole,''
  Astrophys. J. \textbf{875} (2019) no.1, L1
  [arXiv:1906.11238 [astro-ph.GA]].

\bibitem{Nojiri:2006ri}
S.~Nojiri and S.~D.~Odintsov,
``Introduction to modified gravity and gravitational alternative for dark energy,''
eConf \textbf{C0602061}, 06 (2006)
[arXiv:hep-th/0601213 [hep-th]].


\bibitem{Clifton:2011jh}
T.~Clifton, P.~G.~Ferreira, A.~Padilla and C.~Skordis,
``Modified Gravity and Cosmology,''
Phys. Rept. \textbf{513}, 1-189 (2012)
[arXiv:1106.2476 [astro-ph.CO]].


\bibitem{MG3}
  E.~Berti {\it et al.},
  ``Testing General Relativity with Present and Future Astrophysical Observations,''
  Class.\ Quant.\ Grav.\  {\bf 32}, 243001 (2015)
  [arXiv:1501.07274 [gr-qc]].

\bibitem{Fujii}
 Y. Fujii, K. Maeda, "The scalar-tensor theory of gravitation" (Cambridge University Press, 2007).

\bibitem{BBMB}
  N. Bocharova, K. Bronnikov and V. Melnikov, Vestn. Mosk.
  Univ. Fiz. Astron. \textbf{6}, 706 (1970);\\
  J.~D.~Bekenstein, ``Exact solutions of Einstein conformal scalar equations,''
  Annals Phys.\  {\bf 82}, 535 (1974);\\
  J.~D.~Bekenstein, ``Black Holes With Scalar Charge,''
  Annals Phys.\ \textbf{91}, 75 (1975). 

\bibitem{bronnikov}
  K.~A.~Bronnikov and Y.~N.~Kireev,
  ``Instability of Black Holes with Scalar Charge,''
  Phys.\ Lett.\ A {\bf 67}, 95 (1978).

\bibitem{Martinez:1996gn}
  C.~Martinez and J.~Zanelli,
  ``Conformally dressed black hole in (2+1)-dimensions,''
  Phys. Rev. D \textbf{54}, 3830-3833 (1996)
  [arXiv:gr-qc/9604021 [gr-qc]].

\bibitem{Banados:1992wn}
  M.~Banados, C.~Teitelboim and J.~Zanelli,
  ``The Black hole in three-dimensional space-time,''
  Phys. Rev. Lett. \textbf{69}, 1849-1851 (1992)
  [arXiv:hep-th/9204099 [hep-th]].


\bibitem{Martinez:2004nb}
  C.~Martinez, R.~Troncoso and J.~Zanelli,
  ``Exact black hole solution with a minimally coupled scalar field,''
  Phys. Rev. D \textbf{70}, 084035 (2004)
  [arXiv:hep-th/0406111 [hep-th]].


\bibitem{Zloshchastiev:2004ny}
  K.~G.~Zloshchastiev,
  ``On co-existence of black holes and scalar field,''
  Phys.\ Rev.\ Lett.\  {\bf 94}, 121101 (2005)
  [hep-th/0408163].

\bibitem{Martinez:2002ru}
  C.~Martinez, R.~Troncoso and J.~Zanelli,
  ``De Sitter black hole with a conformally coupled scalar field in four-dimensions,''
  Phys.\ Rev.\ D {\bf 67}, 024008 (2003)
  [hep-th/0205319].

\bibitem{Dotti:2007cp}
  G.~Dotti, R.~J.~Gleiser and C.~Martinez,
  ``Static black hole solutions with a self interacting conformally coupled
  scalar field,''
  Phys.\ Rev.\  D {\bf 77}, 104035 (2008)
  [arXiv:0710.1735 [hep-th]].


\bibitem{Torii:2001pg}
  T.~Torii, K.~Maeda and M.~Narita,
  ``Scalar hair on the black hole in asymptotically anti-de Sitter space-time,''
  Phys.\ Rev.\ D {\bf 64}, 044007 (2001).

\bibitem{Winstanley:2002jt}
  E.~Winstanley,
  ``On the existence of conformally coupled scalar field hair for black  holes
  in (anti-)de Sitter space,''
  Found.\ Phys.\  {\bf 33}, 111 (2003)
  [arXiv:gr-qc/0205092].

\bibitem{Martinez:2006an}
  C.~Martinez and R.~Troncoso,
  ``Electrically charged black hole with scalar hair,''
  Phys.\ Rev.\ D {\bf 74}, 064007 (2006)
  [hep-th/0606130].

\bibitem{Kolyvaris:2009pc}
  T.~Kolyvaris, G.~Koutsoumbas, E.~Papantonopoulos and G.~Siopsis,
  ``A New Class of Exact Hairy Black Hole Solutions,''
  Gen.\ Rel.\ Grav.\  {\bf 43}, 163 (2011)
  [arXiv:0911.1711 [hep-th]].

\bibitem{Charmousis:2014zaa}
C.~Charmousis, T.~Kolyvaris, E.~Papantonopoulos and M.~Tsoukalas,
``Black Holes in Bi-scalar Extensions of Horndeski Theories,''
JHEP \textbf{07}, 085 (2014)
[arXiv:1404.1024 [gr-qc]].

\bibitem{Khodadi:2020jij}
  M.~Khodadi, A.~Allahyari, S.~Vagnozzi and D.~F.~Mota,
  ``Black holes with scalar hair in light of the Event Horizon Telescope,''
  JCAP {\bf 2009}, 026 (2020)
  [arXiv:2005.05992 [gr-qc]].

\bibitem{stelle}
 K.~S.~Stelle,
  ``Renormalization of Higher Derivative Quantum Gravity,''
  Phys.\ Rev.\ D {\bf 16}, 953 (1977).

\bibitem{stringT}
  J. Polchinski, String Theory, Vol.$1\&2$ (Cambridge University Press, Cambridge, 2001).

\bibitem{Horndeski}
  G.~W.~Horndeski,
  ``Second-order scalar-tensor field equations in a four-dimensional space,''
  Int.\ J.\ Theor.\ Phys.\  {\bf 10}, 363 (1974).

\bibitem{Deffayet:2009mn}
  C.~Deffayet, S.~Deser and G.~Esposito-Farese,
  ``Generalized Galileons: All scalar models whose curved background extensions maintain second-order field equations and stress-tensors,''
  Phys.\ Rev.\ D {\bf 80}, 064015 (2009)
  [arXiv:0906.1967 [gr-qc]].

\bibitem{Mignemi_1993}
  S.~Mignemi and N.~R.~Stewart,
  ``Charged black holes in effective string theory,''
  Phys.\ Rev.\ D {\bf 47}, 5259 (1993)
  [hep-th/9212146].

\bibitem{Kanti_1996}
  P.~Kanti, N.~E.~Mavromatos, J.~Rizos, K.~Tamvakis and E.~Winstanley,
  ``Dilatonic black holes in higher curvature string gravity,''
  Phys.\ Rev.\ D {\bf 54}, 5049 (1996)
  [hep-th/9511071].

\bibitem{Torrii_1996}
  T.~Torii, H.~Yajima and K.~i.~Maeda,
  ``Dilatonic black holes with Gauss-Bonnet term,''
  Phys.\ Rev.\ D {\bf 55}, 739 (1997)
  [gr-qc/9606034].


\bibitem{Ayzenberg_2014}
  D.~Ayzenberg and N.~Yunes,
  ``Slowly-Rotating Black Holes in Einstein-Dilaton-Gauss-Bonnet Gravity: Quadratic Order in Spin Solutions,''
  Phys.\ Rev.\ D {\bf 90}, 044066 (2014)
  Erratum: [Phys.\ Rev.\ D {\bf 91}, no. 6, 069905 (2015)]
  [arXiv:1405.2133 [gr-qc]].


\bibitem{Kleihaus_2011}
  B.~Kleihaus, J.~Kunz and E.~Radu,
  ``Rotating Black Holes in Dilatonic Einstein-Gauss-Bonnet Theory,''
  Phys.\ Rev.\ Lett.\  {\bf 106}, 151104 (2011)
  [arXiv:1101.2868 [gr-qc]].

\bibitem{Kleihaus_2016a}
  B.~Kleihaus, J.~Kunz, S.~Mojica and M.~Zagermann,
  ``Rapidly Rotating Neutron Stars in Dilatonic Einstein-Gauss-Bonnet Theory,''
  Phys.\ Rev.\ D {\bf 93}, no. 6, 064077 (2016)
  [arXiv:1601.05583 [gr-qc]].

  \bibitem{Bekenstein}
  J.~D.~Bekenstein,
  ``Novel ‘‘no-scalar-hair’’ theorem for black holes,''
  Phys.\ Rev.\ D {\bf 51}, no. 12, R6608 (1995).

\bibitem{Antoniou_2018}
  G.~Antoniou, A.~Bakopoulos and P.~Kanti,
  ``Evasion of No-Hair Theorems and Novel Black-Hole Solutions in Gauss-Bonnet Theories,''
  Phys.\ Rev.\ Lett.\  {\bf 120}, no. 13, 131102 (2018)
  [arXiv:1711.03390 [hep-th]].

\bibitem{Antoniou:2017hxj}
  G.~Antoniou, A.~Bakopoulos and P.~Kanti,
  ``Black-Hole Solutions with Scalar Hair in Einstein-Scalar-Gauss-Bonnet Theories,''
  Phys. Rev. D \textbf{97} (2018) no.8, 084037
  [arXiv:1711.07431 [hep-th]].

\bibitem{Doneva:2017bvd}
  D.~D.~Doneva and S.~S.~Yazadjiev,
  ``New Gauss-Bonnet Black Holes with Curvature-Induced Scalarization in Extended Scalar-Tensor Theories,''
  Phys. Rev. Lett. \textbf{120} (2018) no.13, 131103
  [arXiv:1711.01187 [gr-qc]].

\bibitem{Doneva:2018rou}
  D.~D.~Doneva, S.~Kiorpelidi, P.~G.~Nedkova, E.~Papantonopoulos and S.~S.~Yazadjiev,
  ``Charged Gauss-Bonnet black holes with curvature induced scalarization in the extended scalar-tensor theories,''
  Phys. Rev. D \textbf{98}, no.10, 104056 (2018)
  [arXiv:1809.00844 [gr-qc]].

\bibitem{Herdeiro:2018wub}
  C.~A.~R.~Herdeiro, E.~Radu, N.~Sanchis-Gual and J.~A.~Font,
  ``Spontaneous Scalarization of Charged Black Holes,''
  Phys. Rev. Lett. \textbf{121} (2018) no.10, 101102
  [arXiv:1806.05190 [gr-qc]].

\bibitem{Fernandes:2019rez}
  P.~G.~S.~Fernandes, C.~A.~R.~Herdeiro, A.~M.~Pombo, E.~Radu and N.~Sanchis-Gual,
  ``Spontaneous Scalarisation of Charged Black Holes: Coupling Dependence and Dynamical Features,''
  Class. Quant. Grav. \textbf{36} (2019) no.13, 134002
  [erratum: Class. Quant. Grav. \textbf{37} (2020) no.4, 049501]
  [arXiv:1902.05079 [gr-qc]].

\bibitem{Herdeiro:2019yjy}
  C.~A.~R.~Herdeiro and E.~Radu,
  ``Black hole scalarization from the breakdown of scale invariance,''
  Phys. Rev. D \textbf{99} (2019) no.8, 084039
  [arXiv:1901.02953 [gr-qc]].

\bibitem{Brihaye:2018bgc}
  Y.~Brihaye, C.~Herdeiro and E.~Radu,
  ``The scalarised Schwarzschild-NUT spacetime,''
  Phys. Lett. B \textbf{788} (2019), 295-301
  [arXiv:1810.09560 [gr-qc]].

\bibitem{Silva:2017uqg}
  H.~O.~Silva, J.~Sakstein, L.~Gualtieri, T.~P.~Sotiriou and E.~Berti,
  ``Spontaneous scalarization of black holes and compact stars from a Gauss-Bonnet coupling,''
  Phys. Rev. Lett. \textbf{120} (2018) no.13, 131104
  [arXiv:1711.02080 [gr-qc]].

\bibitem{Minamitsuji:2018xde}
  M.~Minamitsuji and T.~Ikeda,
  ``Scalarized black holes in the presence of the coupling to Gauss-Bonnet gravity,''
  Phys. Rev. D \textbf{99} (2019) no.4, 044017
  [arXiv:1812.03551 [gr-qc]].

\bibitem{Silva:2018qhn}
  H.~O.~Silva, C.~F.~B.~Macedo, T.~P.~Sotiriou, L.~Gualtieri, J.~Sakstein and E.~Berti,
  ``Stability of scalarized black hole solutions in scalar-Gauss-Bonnet gravity,''
  Phys. Rev. D \textbf{99} (2019) no.6, 064011
  [arXiv:1812.05590 [gr-qc]].


\bibitem{Andreou:2019ikc}
  N.~Andreou, N.~Franchini, G.~Ventagli and T.~P.~Sotiriou,
  ``Spontaneous scalarization in generalised scalar-tensor theory,''
  Phys. Rev. D \textbf{99} (2019) no.12, 124022
  [erratum: Phys. Rev. D \textbf{101} (2020) no.10, 109903]
  [arXiv:1904.06365 [gr-qc]].

\bibitem{Minamitsuji:2019iwp}
  M.~Minamitsuji and T.~Ikeda,
  ``Spontaneous scalarization of black holes in the Horndeski theory,''
  Phys. Rev. D \textbf{99} (2019) no.10, 104069
  [arXiv:1904.06572 [gr-qc]].

\bibitem{Peng:2020znl}
  Y.~Peng,
  ``Spontaneous scalarization of Gauss-Bonnet black holes surrounded by massive scalar fields,''
  Phys. Lett. B \textbf{807} (2020), 135569
  [arXiv:2004.12566 [gr-qc]].


\bibitem{Liu:2020yqa}
H.~S.~Liu, H.~Lu, Z.~Y.~Tang and B.~Wang,
``Black hole scalarization in Gauss-Bonnet extended Starobinsky gravity,''
Phys. Rev. D \textbf{103} (2021) no.8, 084043
[arXiv:2004.14395 [gr-qc]].

\bibitem{Doneva:2020qww}
  D.~D.~Doneva, K.~V.~Staykov, S.~S.~Yazadjiev and R.~Z.~Zheleva,
  ``Multiscalar Gauss-Bonnet gravity: Hairy black holes and scalarization,''
  Phys. Rev. D \textbf{102} (2020) no.6, 064042
  [arXiv:2006.11515 [gr-qc]].

\bibitem{Astefanesei:2020qxk}
  D.~Astefanesei, C.~Herdeiro, J.~Oliveira and E.~Radu,
  ``Higher dimensional black hole scalarization,''
  JHEP \textbf{09} (2020), 186
  [arXiv:2007.04153 [gr-qc]].

[53]
\bibitem{Canate:2020kla}
P.~Ca\~nate and S.~E.~Perez Bergliaffa,
``Novel exact magnetic black hole solution in four-dimensional extended scalar-tensor-Gauss-Bonnet theory,''
Phys. Rev. D \textbf{102} (2020) no.10, 104038
[arXiv:2010.04858 [gr-qc]].

\bibitem{Hunter:2020wkd}
  C.~L.~Hunter and D.~J.~Smith,
  ``Novel Hairy Black Hole Solutions in Einstein-Maxwell-Gauss-Bonnet-Scalar Theory,''
  [arXiv:2010.10312 [gr-qc]].


\bibitem{Bakopoulos:2018nui}
  A.~Bakopoulos, G.~Antoniou and P.~Kanti,
  ``Novel Black-Hole Solutions in Einstein-Scalar-Gauss-Bonnet Theories with a Cosmological Constant,''
  Phys. Rev. D \textbf{99} (2019) no.6, 064003
  [arXiv:1812.06941 [hep-th]].

\bibitem{Brihaye:2019gla}
  Y.~Brihaye, C.~Herdeiro and E.~Radu,
  ``Black Hole Spontaneous Scalarisation with a Positive Cosmological Constant,''
  Phys. Lett. B \textbf{802} (2020), 135269
  [arXiv:1910.05286 [gr-qc]].

\bibitem{Bakopoulos:2019tvc}
  A.~Bakopoulos, P.~Kanti and N.~Pappas,
  ``Existence of solutions with a horizon in pure scalar-Gauss-Bonnet theories,''
  Phys. Rev. D \textbf{101} (2020) no.4, 044026
  [arXiv:1910.14637 [hep-th]].

\bibitem{Bakopoulos:2020dfg}
  A.~Bakopoulos, P.~Kanti and N.~Pappas,
  ``Large and ultracompact Gauss-Bonnet black holes with a self-interacting scalar field,''
  Phys. Rev. D \textbf{101} (2020) no.8, 084059
  [arXiv:2003.02473 [hep-th]].

\bibitem{Lin:2020asf}
  K.~Lin, S.~Zhang, C.~Zhang, X.~Zhao, B.~Wang and A.~Wang,
  ``No static regular black holes in Einstein-complex-scalar-Gauss-Bonnet gravity,''
  Phys. Rev. D \textbf{102} (2020) no.2, 024034
  [arXiv:2004.04773 [gr-qc]].

\bibitem{Brihaye:2019dck}
  Y.~Brihaye, B.~Hartmann, N.~P.~Aprile and J.~Urrestilla,
  ``Scalarization of asymptotically anti\textendash{}de Sitter black holes with applications to holographic phase transitions,''
  Phys. Rev. D \textbf{101} (2020) no.12, 124016
  [arXiv:1911.01950 [gr-qc]].

\bibitem{Guo:2020sdu}
  H.~Guo, S.~Kiorpelidi, X.~M.~Kuang, E.~Papantonopoulos, B.~Wang and J.~P.~Wu,
  ``Spontaneous holographic scalarization of black holes in Einstein-scalar-Gauss-Bonnet theories,''
  Phys.\ Rev.\ D {\bf 102}, no. 8, 084029 (2020)
  [arXiv:2006.10659 [hep-th]].

\bibitem{Tang:2020sjs}
Z.~Y.~Tang, B.~Wang, T.~Karakasis and E.~Papantonopoulos,
``Curvature scalarization of black holes in f(R) gravity,''
Phys. Rev. D \textbf{104} (2021) no.6, 064017
[arXiv:2008.13318 [gr-qc]].


\bibitem{Collodel:2019kkx}
  L.~G.~Collodel, B.~Kleihaus, J.~Kunz and E.~Berti,
  ``Spinning and excited black holes in Einstein-scalar-Gauss\textendash{}Bonnet theory,''
  Class. Quant. Grav. \textbf{37} (2020) no.7, 075018
  [arXiv:1912.05382 [gr-qc]].

\bibitem{Dima:2020yac}
A.~Dima, E.~Barausse, N.~Franchini and T.~P.~Sotiriou,
``Spin-induced black hole spontaneous scalarization,''
Phys. Rev. Lett. \textbf{125} (2020) no.23, 231101
[arXiv:2006.03095 [gr-qc]].


\bibitem{Doneva:2020kfv}
D.~D.~Doneva, L.~G.~Collodel, C.~J.~Kr\"uger and S.~S.~Yazadjiev,
``Spin-induced scalarization of Kerr black holes with a massive scalar field,''
Eur. Phys. J. C \textbf{80} (2020) no.12, 1205
[arXiv:2009.03774 [gr-qc]].

\bibitem{Herdeiro:2020wei}
C.~A.~R.~Herdeiro, E.~Radu, H.~O.~Silva, T.~P.~Sotiriou and N.~Yunes,
``Spin-induced scalarized black holes,''
Phys. Rev. Lett. \textbf{126} (2021) no.1, 011103
[arXiv:2009.03904 [gr-qc]].

\bibitem{Berti:2020kgk}
E.~Berti, L.~G.~Collodel, B.~Kleihaus and J.~Kunz,
``Spin-induced black-hole scalarization in Einstein-scalar-Gauss-Bonnet theory,''
Phys. Rev. Lett. \textbf{126} (2021) no.1, 011104
[arXiv:2009.03905 [gr-qc]].

\bibitem{Zhang:2020pko}
S.~J.~Zhang, B.~Wang, A.~Wang and J.~F.~Saavedra,
``Object picture of scalar field perturbation on Kerr black hole in scalar-Einstein-Gauss-Bonnet theory,''
Phys. Rev. D \textbf{102} (2020) no.12, 124056
[arXiv:2010.05092 [gr-qc]].

\bibitem{Motohashi:2018mql}
  H.~Motohashi and S.~Mukohyama,
  ``Shape dependence of spontaneous scalarization,''
  Phys. Rev. D \textbf{99} (2019) no.4, 044030
  [arXiv:1810.12691 [gr-qc]].

\bibitem{Koutsoumbas:2008pw}
G.~Koutsoumbas, E.~Papantonopoulos and G.~Siopsis,
``Phase Transitions in Charged Topological-AdS Black Holes,''
JHEP \textbf{05}, 107 (2008)
[arXiv:0801.4921 [hep-th]].

\bibitem{Wang:2001tk}
B.~Wang, E.~Abdalla and R.~B.~Mann,
``Scalar wave propagation in topological black hole backgrounds,''
Phys. Rev. D \textbf{65}, 084006 (2002)
[arXiv:hep-th/0107243 [hep-th]].

\bibitem{Kolyvaris:2013zfa}
T.~Kolyvaris, G.~Koutsoumbas, E.~Papantonopoulos and G.~Siopsis,
``Phase Transition to a Hairy Black Hole in Asymptotically Flat Spacetime,''
JHEP \textbf{11} (2013), 133
[arXiv:1308.5280 [hep-th]].

\bibitem{Buell} W. Buell and B. Shadwick, Am. J. Phys. 63, 256 (1995).

%
%




\bibitem{BFbound}P. Breitenlohner and D. Z. Freedman, ``Stability in Gauged Extended Supergravity", Annals Phys. 144
(1982) 249.



\bibitem{Gubser1} S. S. Gubser, ``Phase transitions near black hole horizons," Class. Quant. Grav. 22, 5121 (2005) [hep-th/0505189].

\bibitem{Gubser2}S. S. Gubser, ``Breaking an Abelian gauge symmetry near a black hole horizon," Phys. Rev. D 78, 065034 (2008)
[arXiv:0801.2977 [hep-th]].

\bibitem{mann}
  R.~B.~Mann,
  ``Pair production of topological anti-de Sitter black holes,''
  Class.\ Quant.\ Grav.\  {\bf 14}, L109 (1997)
  [arXiv:gr-qc/9607071];\\
  R.~B.~Mann,
  ``Charged topological black hole pair creation,''
  Nucl.\ Phys.\ B {\bf 516}, 357 (1998)
  [arXiv:hep-th/9705223].

\bibitem{Vanzo:1997gw}
  L.~Vanzo,
  ``Black holes with unusual topology,''
  Phys.\ Rev.\ D {\bf 56}, 6475 (1997)
  [arXiv:gr-qc/9705004].


\bibitem{Lemos}
  J.~P.~S.~Lemos and V.~T.~Zanchin,
  ``Rotating Charged Black String and Three Dimensional Black Holes,''
  Phys.\ Rev.\  D {\bf 54}, 3840 (1996)
  [arXiv:hep-th/9511188];\\
  J.~P.~S.~Lemos,
  ``Cylindrical black hole in general relativity,''
  Phys.\ Lett.\  B {\bf 353}, 46 (1995)
  [arXiv:gr-qc/9404041].

\bibitem{Brill:1997mf}
  D.~R.~Brill, J.~Louko and P.~Peldan,
  ``Thermodynamics of (3+1)-dimensional black holes with toroidal or higher
  genus horizons,''
  Phys.\ Rev.\ D {\bf 56}, 3600 (1997)
  [arXiv:gr-qc/9705012].

\bibitem{Birmingham}
  D.~Birmingham,
  ``Topological black holes in anti-de Sitter space,''
  Class.\ Quant.\ Grav.\  {\bf 16}, 1197 (1999)
  [arXiv:hep-th/9808032].

\bibitem{Cai:1996eg}
R.~G.~Cai and Y.~Z.~Zhang,
``Black plane solutions in four-dimensional space-times,''
Phys. Rev. D \textbf{54}, 4891-4898 (1996)
[arXiv:gr-qc/9609065 [gr-qc]].

\bibitem{Huang:1995zb}
C.~G.~Huang and C.~B.~Liang,
``A Torus like black hole,''
Phys. Lett. A \textbf{201}, 27-32 (1995)

\bibitem{Cai:2001dz}
R.~G.~Cai,
``Gauss-Bonnet black holes in AdS spaces,''
Phys. Rev. D \textbf{65}, 084014 (2002)
[arXiv:hep-th/0109133 [hep-th]].

\bibitem{Gibbons:2002pq}
  G.~Gibbons and S.~A.~Hartnoll,
  ``A gravitational instability in higher dimensions,''
  Phys.\ Rev.\  D {\bf 66}, 064024 (2002)
  [arXiv:hep-th/0206202].

\bibitem{Birmingham:2007yv}
  D.~Birmingham and S.~Mokhtari,
  ``Stability of Topological Black Holes,''
  Phys.\ Rev.\  D {\bf 76}, 124039 (2007)
  [arXiv:0709.2388 [hep-th]].

\bibitem{Koutsoumbas:2006xj}
  G.~Koutsoumbas, S.~Musiri, E.~Papantonopoulos and G.~Siopsis,
  ``Quasi-normal Modes of Electromagnetic Perturbations of Four-Dimensional Topological Black Holes with Scalar Hair,''
  JHEP \textbf{10}, 006 (2006)
  [arXiv:hep-th/0606096 [hep-th]].


\bibitem{Koutsoumbas:2008yq}
  G.~Koutsoumbas, E.~Papantonopoulos and G.~Siopsis,
  ``Discontinuities in Scalar Perturbations of Topological Black Holes,''
  Class. Quant. Grav. \textbf{26}, 105004 (2009)
  [arXiv:0806.1452 [hep-th]].

%

\end{thebibliography}

\end{document}